\documentclass[usenatbib]{mnras}

\usepackage[utf8]{inputenc}
\usepackage{amssymb}
\usepackage{amsmath}
\usepackage{relsize}
\usepackage[pdftex]{graphicx}
\usepackage{todonotes}
\usepackage{cleveref} %multiple references 
\usepackage{mathptmx}
\usepackage{color}
\usepackage{pdflscape}
\usepackage{siunitx}
\usepackage{multirow}
\usepackage{multicol}
\usepackage{makecell}

\newcommand{\be}{\begin{equation}}
\newcommand{\ee}{\end{equation}}

\makeatletter
\newcommand*{\rom}[1]{\expandafter\@slowromancap\romannumeral #1@}
\makeatother

\newcommand{\pd}[2]{\frac{\partial #1}{\partial #2}}

\newcommand*\hi{{\rm{H}\,\textsc{i}}}

\newcommand*\HI{H\,{\textsc{i}}}

\newcommand*\diff{\mathop{}\!\mathrm{d}}

\begin{document}
\title{Intensity Mapping as a Probe of axion dark matter}
\author[J. B. Bauer et al.] {Jurek~B.~Bauer$^1$\thanks{jurek.bauer@uni-goettingen.de}, David~J.~E.~Marsh$^1$\thanks{david.marsh@uni-goettingen.de}, Ren\'{e}e Hlo\v{z}ek$^2$,
\newauthor
Hamsa Padmanabhan$^{2,3,4}$, Alex Lagu\"{e}$^{2,3}$\\
$^1$Institut f\"{u}r Astrophysik, Georg-August-Universit\"{a}t, Friedrich-Hund-Platz 1, D-37077 G\"{o}ttingen, Germany. \\
$^2$Dunlap Institute for Astronomy and Astrophysics and Department of Astronomy and Astrophysics\\ University of Toronto, 50 St George Street, Toronto, ON M5S 3H4, Canada. \\
$^3$Canadian Institute for Theoretical Astrophysics, 60 St George Street, Toronto, ON M5S 3H8, Canada.\\
$^4$D\'{e}partement de Physique Th\'{e}orique, Universit\'{e} de Gen\`eve, 24 Quai Ernest-Ansermet, CH-1211 Gen\`eve 4, Switzerland.\\
}

\date{\today}
\maketitle

\begin{abstract}
We consider intensity mapping (IM) of neutral hydrogen (\HI) in the redshift range $0 \lesssim z \lesssim 3$ employing a halo model approach where \HI is assumed to follow the distribution of dark matter (DM) halos. 
If a portion of the DM is composed of ultralight axions, then the abundance of halos is changed compared to cold DM below the axion Jeans mass. 
With fixed total \HI density, $\Omega_{\hi}$, assumed to reside entirely in halos, this effect introduces a scale-independent increase in the \HI power spectrum on scales above the axion Jeans scale, which our model predicts consistent with $N$-body simulations.
Lighter axions introduce a scale-dependent feature even on linear scales due to its suppression of the matter power spectrum near the Jeans scale.
 We use the Fisher matrix formalism to forecast the ability of future \HI surveys to constrain the axion fraction of DM and marginalize over astrophysical and model uncertainties. We find that a HIRAX-like survey is a very reliable IM survey configuration, being affected minimally by uncertainties due to non-linear scales, while the SKA1MID configuration is the most constraining as it is sensitive to non-linear scales.
 Including non-linear scales and combining a SKA1MID-like IM survey with the Simons Observatory CMB, the benchmark ``fuzzy DM'' model with $m_a = 10^{-22}\text{ eV}$ can be constrained at few percent. This is almost an order of magnitude improvement over current limits from the Ly\,$\alpha$ forest.
 For lighter ULAs this limit improves below 1\%, and allows the possibility to test the connection between axion models and the grand unification scale across a wide range of masses.
\end{abstract}

\section{Introduction}
\label{sec:intro} 
Measurements of the power spectrum of the cosmic microwave background (CMB) anisotropies establish the precision cosmological standard model~\citep{2016A&A...594A..11P,2018arXiv180706209P}. Intensity mapping (IM) of spectral lines has great potential as a future cosmological probe~\citep{Loeb_Wyithe2008, Bull_Ferreira2015, intensity_mapping_status_report2017, Kovetz_etal2019, Bernal2019, Padmanabhan_etal_forecast, 2019BAAS...51g.241P}, since the frequency dependence due to redshift gives a tomographic three-dimensional map, vastly increasing the number of accessible modes compared to the CMB~\citep{Mao_Tegmark_etal2008}. 

Hydrogen is the most abundant element in the Universe: According to measurements and the standard theory of big bang nucleosynthesis, it makes up approximately 75\% of all ordinary matter~\citep{2018arXiv180706209P}. In the intergalactic medium at $z\lesssim 6$, hydrogen is ionized by UV radiation from galaxies, while neutral hydrogen (\HI) resides to great extent (about 80--90\%) in comparatively dense clouds, which shields them from ionizing UV radiation. These clouds are known as damped Ly\,$\alpha$ (DLA) systems and are confined to galaxies~\citep{Prochaska_etal2005, Zwaan_etal2005a, Lah_etal2007, Lah_etal2009}. 
Also, simulations have repeatedly found that $>90\%$ of the neutral gas is in halos \citep[see][and references therein]{ingredients_for_21cm_IM}.
 Therefore, \HI traces the structure of galaxies and their host dark matter (DM) halos, and IM of the hyperfine, or 21cm, \HI transition is a probe of DM clustering. An empirical (data-driven) framework for the \HI power spectrum is provided by the \HI halo model~\citep{Refregier_Padmanabhan2017,Refregier_Padmanabhan_Amara2017}, which maps between the theoretical DM halo mass function~\citep[e.g][]{Sheth_Tormen2002}, and the \HI halos that trace it \citep[see e.g.][for other \HI halo prescriptions]{Bagla2010, Marin_etal2010}.

DM is  a key ingredient in the cosmological standard model, yet its nature is a mystery. The only DM candidate in the standard model of particle physics is the neutrino, which is known to make up less than 1\% (but more than 0.5\%)
 of the total DM abundance because the relativistic velocity of the neutrino background make it too ``hot'' to account for observed structure formation~\citep{SDSS2017, 2018arXiv180706209P, ParticleDataGroup_review2018}. 
Observations are consistent with the majority of the remaining DM being composed of a single species of cold, collisionless DM (CDM). Of relevance to this study, CMB anisotropies constrain the density parameter of ultralight axions (ULAs) to be $\Omega_a h^2\lesssim 0.003$ over the mass range $10^{-32} \leq m_a\leq 10^{-25}\text{ eV}$~\citep{Hlozek_Marsh2018}.

The existence of multiple species of light axions is a generic and well-established prediction of string/M-theory~\citep{Conlon:2006tq,Svrcek:2006yi,Acharya:2010zx,Arvanitaki_2010,Demirtas:2018akl} and many other extensions of the standard model~\citep[e.g][]{PecceiQuinn1977,Weinberg1978,Wilczek1978,Banks:2002sd,Kim:2015yna}. The axion density parameter is determined by the axion mass, $m_a$, and symmetry breaking scale (or ``decay constant''), $f_a$, as well as a single random number, $|\theta_i|\in [0,\pi]$, related to the initial conditions~\citep[see][for a review]{Marsh_review2016}. In the mass range of interest to cosmology, the decay constant is constrained to be in the range $10^9 \lesssim f_a\lesssim 10^{18}\text{ GeV}$, with some models preferring the grand unified scale, $f_a\sim 10^{16}\text{ GeV}$. Taking $\theta_i\sim\mathcal{O}(1)$, it is thus predicted that ULAs with $m_a\lesssim 10^{-22}\text{ eV}$ contribute subdominantly to the DM density (consistent with observations). On the other hand those with $m_a\sim 10^{-22}\text{ eV}$, known as ``fuzzy DM''~\citep{Hu:2000ke}, can make up a significant fraction of the DM, and furthermore have a host of interesting phenomenological consequences on galaxy formation and other astrophysical systems~\citep[see e.g.][]{Arvanitaki_2010,Schive:2014dra,Hui:2016ltb,Marsh_review2016,Niemeyer:2019aqm}.

In this work, we use the \HI halo model to explore the effects of a ULA subspecies of DM in the mass range $10^{-32}\leq m_a\leq 10^{-22}\text{ eV}$ on the \HI power spectrum at $z\leq 6$. Due to the large ULA de Broglie wavelength, small-scale structure formation is suppressed relative to CDM, which manifests in an \emph{increase} in the \HI power on large scales (see Section \ref{sec:results_HI_PS}). We show that, due to this effect, future IM surveys in conjunction with the CMB are sensitive to a per-cent-level ULA component of DM, and can thus be used as a precision test of the predictions related to fuzzy DM and the grand unified scale for $f_a$.

This paper is organized as follows: First, we introduce the formalism in Section \ref{sec:formalism}. This includes a description of the axion physics at play (Section \ref{sec:axion_physics}) and the \HI halo model and how axions are accommodated to it (Section \ref{sec:HI_signal}). We present the results on the \HI power spectrum and compare them to those of pure ULA numerical simulations in Section \ref{sec:results_HI_PS}. In Section \ref{sec:experimental_setup}, the configurations for the IM surveys are described, and in Section \ref{sec:fisher}, the Fisher forecast formalism  is introduced. We proceed by presenting the main results and constraints gained within this framework in Section \ref{sec:results} and discuss our main findings in Section \ref{sec:discussion}.

\section{The \HI Power Spectrum}
\label{sec:formalism}
21\,cm IM measures the integrated intensity of the spin-flip transition of neutral hydrogen across the sky and redshift \citep[see e.g.][and references therein]{Bull_Ferreira2015}.
The redshifted \SI{21}{cm} radiation is well into the radio regime which means that matter along the line of sight does not interfere with the signal. The redshifted signal can be detected from the ``Dark Ages'', through the epoch of reionization (EoR) and the post-reionization Universe. 
After reionization the remaining neutral hydrogen is expected to reside to great extent in galactic haloes.
Thus, neutral hydrogen is expected to trace the galaxy distribution in the current, post-reionization epoch ($z \lesssim 6$). 
Over recent years, it became apparent that using IM to measure the large-scale structure in the late-time Universe is a promising cosmological probe \citep{Bharadwaj2001, Chang2010, Battye_2012, Bull_Ferreira2015, weighing_neutrinos2015}.
In this work we especially focus on the prospect of these surveys to constrain cosmologies including ultra-light axions (ULAs).
To do this we employ and modify an empirical framework to describe the 21\,cm signal conceived and further constrained by Padmanabhan et al. \citep{Padmanabhan_etal2015, Padmanabhan_etal2016, Refregier_Padmanabhan2017, Refregier_Padmanabhan_Amara2017}.  
This formalism effectively treats the neutral hydrogen as a biased tracer of the underlying matter distribution
and
the model parameters are entirely constrained by the compilation of the latest observations on neutral hydrogen systems over $z \sim 0-5$.
The astrophysical priors thus obtained are realistic since they are grounded in present-day observations.
The advantages of using such an empirical model are manifold: 
Due to the computational simplicity, it allows us to consider many models, vary physics easily, and it can be conveniently implemented within a Fisher matrix analysis. 
Furthermore, our model reproduces the qualitative features of simulations with ULA DM by \citet{Carucci_2017}. 
\subsection{Axion Physics}
\label{sec:axion_physics}
\begin{table}
\centering
\caption{Fiducial cosmological and astrophysical parameters and step size for calculating the Fisher derivatives. The cosmological parameters are the same as in the forecast by \citet{Hlozek_Marsh_forecast} and are within current CMB data constraints in \citet{Hlozek_Marsh2018}. The astrophysical parameters have been adopted from \citet{Padmanabhan_etal_forecast} and are the best-fitting values found in \citet{Refregier_Padmanabhan_Amara2017}.}
\begin{tabular}{l c c}
\hline
\hline
Parameter & Fiducial value & Step size \\
\hline
$h$ & 0.69 & 0.01 \\
$\Omega_d$ & 0.25142 & 0.004 \\
$\Omega_b$ & 0.04667 & 0.004 \\
$\Omega_a/\Omega_d$ & 0.02 & 0.005 \\
$\sum m_\nu$ $[\si{eV}]$ & 0.06 & fixed \\
$N_{\mathrm{eff}}$ & 3.046 & fixed \\
$A_s$ & $\SI{2.1955e-9}{}$ & $\SI{e-13}{}$ \\
$n_s$ & 0.9655 & 0.0005 \\
$k_{\mathrm{piv}}$ $[\si{Mpc^{-1}}]$ & 0.05 & fixed \\
$m_a$ $[\si{eV}]$ & $10^{-32} < m_a < 10^{-22}$ & fixed per run\\
\hline
$v_{c,0}$ $[\si{km/s}]$ & $36.3$ & $0.01$ \\
$\beta$ & $-0.58$ & $0.003$ \\
$\gamma$ & 1.45 & fixed \\
$c_{\hi,0}$ & 28.65 & fixed \\ 
$\Omega_\hi h$ & $\SI{2.45e-4}{}$ & fixed \\
\hline
\hline
\end{tabular}
\label{tab:parameters}
\end{table}
In this section we shortly recapitulate the relevant linear physics of axions. It is included in {\textsc{axioncamb}}\footnote{Publicly available at \url{https://github.com/dgrin1/axionCAMB}.} \citep{Hlozek_Marsh2015}, which we use to perform the calculations. 
ULAs are described as a pseudo-scalar field obeying the Klein--Gordon equation for temperatures below the global symmetry breaking and non-perturbative scales.
\footnote{In summary, we treat the axions with potential $V=m_a^2 \phi^2/2$, time-independent mass $m_a$, and no interactions, e.g. axion-photon conversion in the presence of a magnetic field \citep[e.g. see][for the impact on IM surveys of a possible two-photon decay of axion]{Creque-Sarbinowski_Kamionkowski2018}.} 
This classical treatment of the axion field is justified due to huge occupation number of a condensate with cosmological density.

Axion DM is produced by vacuum realignment of the classical field \citep{1983PhLB..120..137D, 1983PhLB..120..127P, 1983PhLB..120..133A}.
The axion field at early times in the Universe, shortly after inflation, is overdamped and therefore mimics the vacuum energy with equation-of-state parameter $w=-1$. Later, when $m_a \sim H$ (with $H := \dot{a}/a$ and $a$ being the scale factor in the FLRW metric) the axion field starts to oscillate, defining $a_{\mathrm{osc}}$.
From that time on the energy density scales as $\rho_a \sim a^{-3}$, just as ordinary matter (and the pressure $p_a$ and $w_a=p_a/\rho_a$ oscillate rapidly around zero). 
This makes the axion field a suitable candidate for CDM. 
The axion density parameter depends on the value of the axion energy density at $a_{\mathrm{osc}}$, i.e. $\Omega_a = \rho_a(a_{\mathrm{osc}}) a_{\mathrm{osc}}^3 /\rho_{\mathrm{crit}}$ \citep{Marsh_Ferreira2010}, with $\rho_{\mathrm{crit}}$ being the cosmological critical energy density today.
A useful approximate formula for the axion density parameter is given by \citep{Hlozek_Marsh2015}:
\begin{align}
\Omega_a \approx \begin{cases}
\num{3.3e-3} \left(\frac{\Omega_m}{0.3} \frac{3600}{1+z_\mathrm{eq}}\right)^{3/4} \\
\hspace*{0.5cm} \times \left( \frac{m_a}{\SI{e-22}{eV}}\frac{\SI{69}{\frac{km}{s Mpc}}}{H_0}\right)^{1/2} 
\left(\frac{\phi_i}{\SI{e16}{GeV}}\right)^2, & \text{if } a_{\mathrm{osc}} < a_{\mathrm{eq}} \\
\num{7.6e-6} \left(\frac{\Omega_m}{0.3}\right) \left(\frac{\phi_i}{\SI{e16}{GeV}}\right)^2, & \text{if } a_{\mathrm{eq}} < a_{\mathrm{osc}} < 1.
\end{cases}
\label{eq:axion_abundance}
\end{align}
In the following, we will parametrize the axion abundance relative to the total DM density parameter with $\Omega_a/\Omega_d$ and $\Omega_d = \Omega_c + \Omega_a$.

Perturbation in this axion field can be solved with help of a WKB-like ansatz, once the scalar field is in its oscillary phase. One finds\footnote{Primes denote derivatives with respect to the conformal time.} \citep{Hwang_Noh2009} that $\langle w \rangle = \langle w' \rangle =0$ with a non-negligible sound speed arising from the large de Broglie wavelength of the axion: 
\begin{align}
c_s^2 = \frac{\frac{k^2}{4 m_a^2 a^2}}{1 + \frac{k^2}{4 m_a^2 a^2}}.
\label{eq:sound_speed}
\end{align}
Thus, perturbations in the axion field are subject to a pressure induced by the uncertainty principle (relevant at cosmological scales due to the tiny mass of the axion).
The equations of motion in synchronous gauge for the perturbed axion energy density is that of a general fluid with the above parameters (when $a \gg a_{\mathrm{osc}}$) \citep{Marsh_review2016}
\begin{align}
\label{eq:axion_pert_eom1}
\delta^\prime_{a} &= -k u_{a}-\frac{h^\prime}{2}-3 \mathcal{H} c_{s}^{2} \delta_{a}-9 \mathcal{H}^{2} c_{s}^{2} u_{a} / k, \\
u^\prime_{a} &= -\mathcal{H} u_{a}+c_{s}^{2} k \delta_{a}+3 c_{s}^{2} \mathcal{H} u_{a}.
\label{eq:axion_pert_eom2}
\end{align}
Here, $\mathcal{H}$ denotes the conformal Hubble rate and $u_a$ the dimensionless perturbed heat flux.
The additional non-canonical terms on the right-hand side of equations \eqref{eq:axion_pert_eom1} and \eqref{eq:axion_pert_eom2} account for equation \eqref{eq:sound_speed} applying only in the axion comoving gauge \citep{Hlozek_Marsh2015}. 

From equations \eqref{eq:axion_pert_eom1} and \eqref{eq:axion_pert_eom2}, it is evident that for large scales the pressure terms go to zero, $c_s^2 \rightarrow 0$, and the dynamics of the axion equations of motion match those for CDM. 
For smaller scales, however, the sound speed term becomes relevant, giving rise to a scale-dependent, oscillating solution, ultimately responsible for the suppression of structure when compared to CDM. 
\begin{figure}
\centering
    \includegraphics[width=\linewidth]{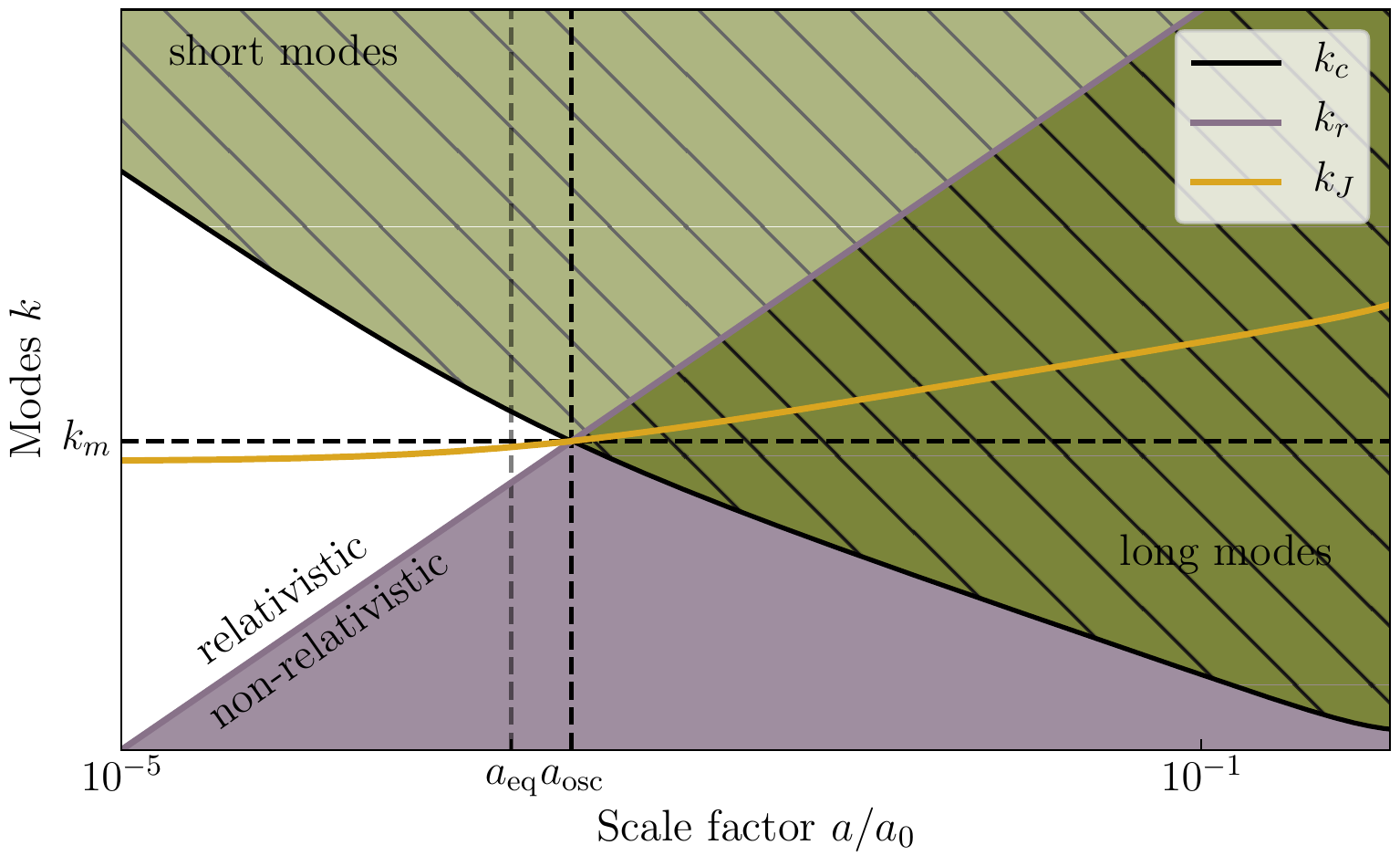}
    \caption{Evolution of the relevant scales for axion perturbations for an axion mass of $m_a = \SI{e-28}{eV}$ and other parameters as given in Table \ref{tab:parameters}. The hatched region shows the scales inside the Hubble horizon (equation \ref{eq:horizon_crossing}) and the purple-shaded region indicates non-relativistic modes (where $c_s^2 \approx k^2/(4m_a^2a^2)$).}
    \label{fig:scales_overview}
\end{figure}

To get an idea how this scale- and time-dependent sound speed term affects the evolution of axion density perturbations, let us first define the scale at horizon crossing, 
\begin{align}
k_c = a H,
\label{eq:horizon_crossing}
\end{align}
and the scale where oscillations roughly start, $k_r \sim m_a a$.
Modes with $k > k_r$ are thus relativistic modes with $c_s^2 \approx 1$, while modes with $k < k_r$ are non-relativistic with $c_s^2 \rightarrow k^2/(4 m_a^2 a^2)$. 
As $k_r$ increases with time, more and more modes become non-relativistic.  

If a mode is already non-relativistic when it enters the horizon ($k_c < k_r \Leftrightarrow H < m_a$), the sound speed term is negligibly small and the mode will behave as ordinary DM (``long modes''). 
If, however, the mode is relativistic when entering the horizon ($m_a > H$) and becomes non-relativistic later on, the sound speed term cannot be neglected and the axions ``free-stream'' (``short modes''). 
These modes will decay until some later time (when the gravitational pull is dominant), given by the comoving Jeans wavenumber $k_J = a \sqrt{H m}$~\citep{Khlopov_etal1985}.
The comoving Jeans wavenumber is time-independent in a radiation-dominated epoch. So, relativistic modes entering the horizon when radiation dominates will decay until after matter--radiation equality.
Let us define the minimal scale $k_m$ at which suppression, i.e. no growth of modes, sets in for given mass $m_a$ (cf. Fig. \ref{fig:scales_overview}). 
The axions therefore introduce a step-like feature to the matter transfer function, which is due to its scale-dependent sound speed term \citep{Arvanitaki_2010}. The width of the step is given by $k_m$ and $k_J$.
The different scales at play are shown in Fig. \ref{fig:scales_overview} for an axion of mass $m_a = \SI{e-28}{eV}$.

Finally, note that this behavior is conceptually similar to that of massive neutrinos. 
Due to their large thermal velocities, massive neutrinos also introduce an effective sound speed term $c_s = T_\nu^0/(m_\nu a)$ for $a \gtrsim T_\nu^0/m_\nu$, where $T_\nu^0$ is the neutrino temperature today and $m_\nu$ the neutrino mass \citep[e.g.][]{Amendola_Barbieri2006, Marsh_etal2012}. 
This, equivalently to $k_m$ and $k_J$ for axions, defines a ``free-streaming'' scale below which the pressure term dominates and clustering is prohibited. 
Analogous to axions, this also introduces a step-like feature to the transfer function, i.e. a suppression of the matter power spectrum above $k_m$ when compared to CDM.

\subsection{Modeling \HI}
\label{sec:HI_signal}
In the following, we introduce the \HI halo model exploited for this study (Section \ref{sec:general_model}) and specifically comment on the inclusion of ULAs in Section \ref{sec:axionsimprint}. 
Halo models are concerned about amplitudes of matter fluctuations larger than one \citep{Cooray_Sheth2002} and matter components with small variance can approximated as not bound within halos.
For axions, the latter statement is mass dependent as we will show in Section \ref{sec:axionsimprint}.
\subsubsection{The Halo Model and Angular Power Spectrum}
\label{sec:general_model}
We exploit spherical harmonic tomography to analyze the two-point correlation of the \HI fluctuation. This choice circumvents the need to assume a specific comoving distance relation $r(z)$ and, therefore, a specific cosmology when analyzing the data.
Spherical harmonic tomography discretizes the redshift range and decomposes the signal in each redshift bin with spherical harmonics. 
The measured brightness temperature $\delta T(\bmath{x}, z)$ is projected on the sky 
with the commonly used projection kernel \citep{Battye_2012}:
\begin{align}
W_i(z) = \begin{cases}
\frac{1}{\Delta z}, & \text{if } z_i - \tfrac{\Delta z}{2} \leq z \leq z_i + \tfrac{\Delta z}{2} \\
0, & \text{otherwise.}
\end{cases}
\end{align}
The dimensionless 2D angular power spectrum (dividing by the mean brightness temperature) is ultimately given by  
\begin{align}
\label{eq:C_ell_without_Limber}
C_\ell(z_i, z_j) = \frac{2}{\pi} \int \diff z\,W_i(z)  \int \diff z'\,W_j(z') \\
\nonumber \times \int k^2 \diff k\,P_\hi(k,z,z') j_\ell(kr(z')) j_\ell(kr(z)),
\end{align}
where $j_\ell(x)$ is the spherical Bessel function of the first kind and $P_\hi(k,z,z')$ denotes the unequal-time \HI power spectrum.\footnote{Note that this quantity is without further considerations not purely expressible in terms of the equal-time \HI power spectrum as defined above \citep[cf. the discussion after equation 18 of][]{Camera_Padmanabhan2019}.} The comoving distance to redshift $z$ is given by (with $c$ being the speed of light)
\begin{align}
r(z) = c \int_0^{z} \frac{\diff z'}{H(z')}.
\end{align}

In the Limber approximation \citep{Limber1953}, the spherical Bessel function of the first kind is approximated by $j_\ell(x)~\rightarrow~\sqrt{\frac{\pi}{2\ell +1}}\delta_D(x - (\ell+\tfrac{1}{2}))$, and correlations between different redshift bins cancel. Thus, we calculate the dimensionless angular power spectrum by:
\begin{align}
\label{eq:dimensionless_Cl_with_Limber}
C_\ell(z_i) \simeq \frac{1}{c} \int \diff z \frac{W_i^2(z) H(z)}{r^2(z)} P_\hi\left(\frac{\ell+\tfrac{1}{2}}{r(z)}, z\right),
\end{align}
where $P_\hi(k,z)$ denotes the Cartesian \HI power spectrum.
\citet{LoVerde_Ashfordi_2008} showed that in the case of narrow redshift bins the approximation is expected to be accurate within 1\% above $\ell \sim 10$ for $P(k,z) = P(k) D^2(z)$. 
In this study, similar narrow redshift bins are used (see below). 
If the above factorization of wavenumber $k$ and redshift $z$ does not hold, second-order corrections to the Limber approximation arise. 
However, this factorization does hold in the redshift range considered presently for ULAs with mass $> \SI{e-32}{eV}$.
Out of this reason and given that ULAs leave the matter power spectrum unchanged on large scales exactly where the Limber approximation is known to be less accurate, we do not expect that the accuracy changes significantly for $C_\ell$ in the present case (except for the marginal case $m_a \sim \SI{e-32}{eV}$). 
Moreover, \citet{Olivari2017} showed that the difference between the exact formula and the Limber approximation is small: The loss in information from the cross-correlation of different redshift bins is roughly compensated in the enhancement of the autocorrelation. 

Note that in deriving \Cref{eq:C_ell_without_Limber,eq:dimensionless_Cl_with_Limber} the effect of peculiar velocities was neglected \citep{Battye_2012}.
The latter will lead to redshift-space distortions (RSDs), which manifest themselves e.g. in the Kaiser effect \citep{Kaiser1987} and the Sachs--Wolfe effects \citep[for a thorough discussion, see e.g.][]{Bonvin_Durrer2011}.
\citet{Seehars_2016} showed by a seminumerical approach (i.e. taking an $N$-body simulation and populating the halos with neutral hydrogen via \HI--halo mass relation) that the RSDs may be significant for large scales ($\ell \lesssim 200$). 
None the less, we choose to neglect the RSDs in this paper.
Ignoring the RSDs is a conservative assumption in that including them would provide additional cosmological constraints \citep{Bull_Ferreira2015} and  increase the \SI{21}{cm} signal \citep[cf.][]{Seehars_2016}. 
Likewise, including RSDs leads to significantly larger theoretical uncertainties, due to the complexity of modelling non-linear scales.

It  should be  noted  that  they  also  increase  the  complexity  of  modelling  non-linear  scales, leading  to significantly larger theoretical uncertainties.

\begin{figure}
\centering
    \includegraphics[trim={0.5cm 0.5cm 0.7cm 0.5cm},clip,width=1.\linewidth]{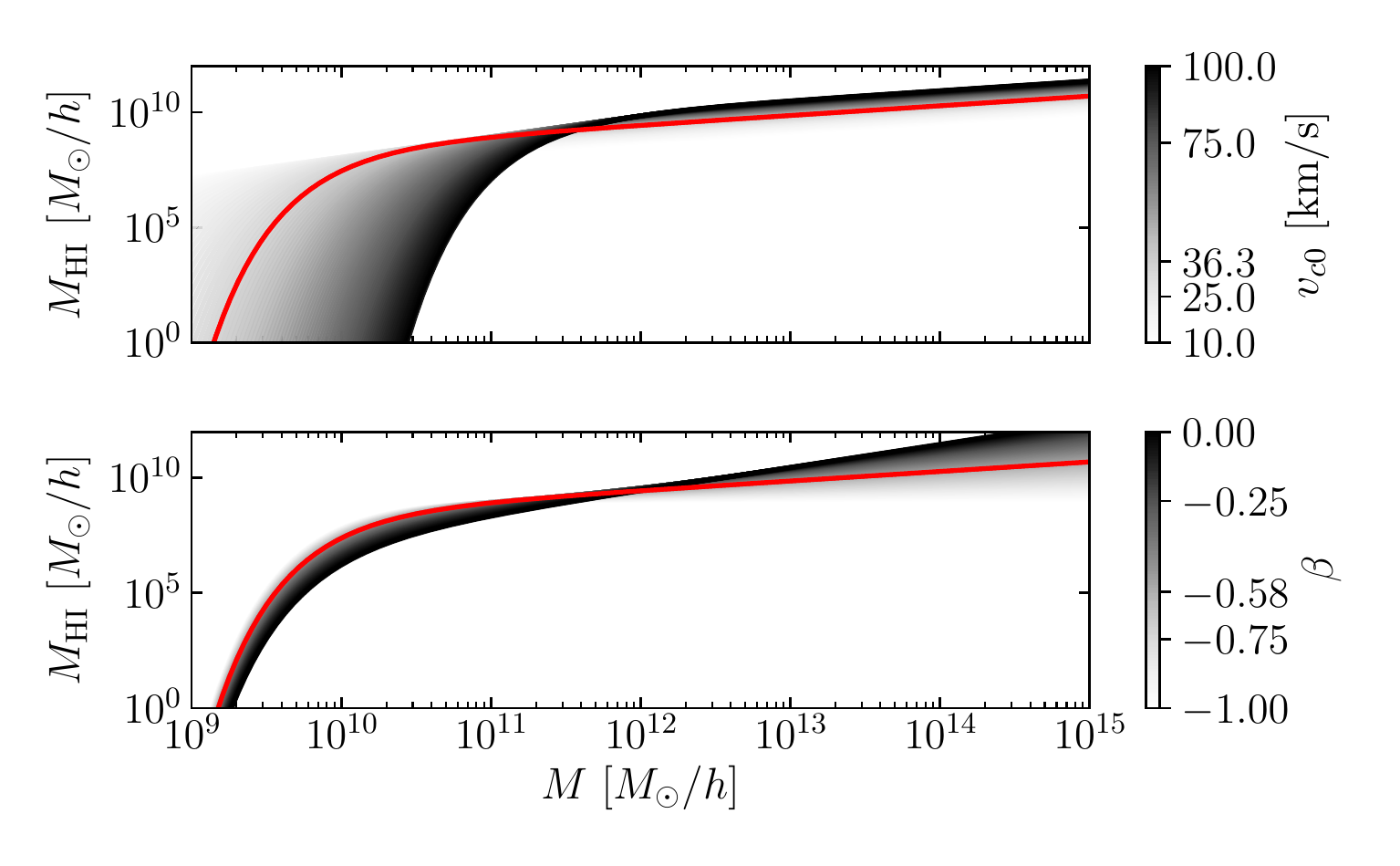}
    \caption{The \HI--halo mass relation from \citet{Refregier_Padmanabhan_Amara2017}. The normalization $\alpha$ was fixed to match the neutral hydrogen density for a cosmology with $m_a = \SI{e-24}{eV}$ and other cosmological parameters given in Table \ref{tab:parameters} at redshift 0. Red lines denote the fiducial values from the best-fitting model in \citet{Refregier_Padmanabhan_Amara2017} and the grey-shaded regions show the effect upon varying the \HI model parameters.}
    \label{fig:M_HI}
\end{figure}
To calculate the \HI power spectrum we exploit the neutral hydrogen halo model, elaborated on in several papers \citep{Padmanabhan_etal2016, Refregier_Padmanabhan2017, Refregier_Padmanabhan_Amara2017}. The authors constrained it with the combination of all existing low-redshift \SI{21}{cm} observations, DLA data and \HI galaxy surveys \citep[e.g.][respectively]{Switzer_etal2013, Zafar_etal2013, Martin_etal2012}. 
Specifically, the halo model assumes that a halo of mass $M$ at redshift $z$ contains a number of galaxies that, in total, host neutral hydrogen of mass $M_\hi$. 
Compactly, this statement is condensed in the deterministic function, 
\begin{align}
M_\hi(M, z) = \alpha f_{\mathrm{H},c} M \left(\frac{M}{10^{11} h^{-1} M_\odot}\right)^\beta \exp\left[-\left(\frac{v_{c0}}{v_c(M, z)}\right)^3\right].
\label{eq:HI_halo_mass_relation}
\end{align}
This relation features three free parameters. $\alpha$ is the overall normalization, $\beta$ is the slope of the \HI--halo mass relation, and $v_{c,0}$ is the physical, lower virial velocity cut-off, above which a halo can host neutral hydrogen.
$f_{\mathrm{H},c}$ is the cosmic hydrogen fraction: $f_{\mathrm{H},c} = \Omega_b/\Omega_m(1 - Y_p)$, where $Y_p=0.24$ is the helium abundance.

The virial velocity of the halo, $v_c(M,z)$, is related to its mass $M$ by
\begin{align}
v_c(M,z) = \sqrt{\frac{G M}{R_{\mathrm{vir}}}}, 
\label{eq:virial_velocity}
\end{align}  
where the virial radius in physical coordinates is
\begin{align}
R_{\mathrm{vir}} = \left(\frac{3}{4 \pi}\frac{M}{\Delta_{\mathrm{h}} \rho_0}\right)^{1/3} \frac{1}{1 + z}.
\label{eq:virial_radius}
\end{align} 
$\rho_0 = \Omega_0 \rho_{\mathrm{crit}}$ is the background density, also used to calculate the halo mass function and bias (this choice will be discussed in Section \ref{sec:axionsimprint}). 
The virial parameter $\Delta_{\mathrm{h}}$ takes the following form \citep{Bryan_Norman1998}
\begin{align}
\Delta_{\mathrm{h}} = 18 \pi^2 + 82 d - 39d^2,\\
d := \frac{\Omega_m (1+z)^3}{E(z)^2} - 1,
\end{align}
where $E(z)$ is given by $H(z) = H_0 E(z)$.
The \HI--halo mass relation is shown in Fig. \ref{fig:M_HI}, where the normalization $\alpha$ was fixed to match the neutral hydrogen density for a cosmology with $m_a = \SI{e-24}{eV}$ and other cosmological parameters as given in Table \ref{tab:parameters}.

Furthermore, to make fully use of the halo model an exponential profile for the \HI is assumed, which is well motivated and commonly used \citep[e.g.][]{Binney_Tremaine_book, Bluedisk_survey}:
\begin{align}
\varrho_\hi(r,M) = \varrho_{\hi,0} {\rm e}^{-r/r_s},
\end{align}
where $r_s = R_{\mathrm{vir}}(M, z)/c_\hi(M,z)$. $\varrho_{\hi,0}$ is fixed such that integrating the radial profile over a sphere of radius $R_{\mathrm{vir}}$ matches $M_\hi(M,z)$.
$R_{\mathrm{vir}}$ is the virial radius given in equation \eqref{eq:virial_radius} and $c_\hi(M,z)$ is the \HI concentration parameter given by \citep{Refregier_Padmanabhan_Amara2017}
\begin{align}
c_\hi(M,z) = c_{\hi,0} \left(\frac{M}{\SI{e11}{M_\odot}}\right)^{-0.109} \frac{4}{(1+z)^\gamma}.
\label{eq:HI_concentration}
\end{align}
The radial profile, therefore, has two free parameters, $c_{\hi,0}$ and $\gamma$. 
\HI IM constrains these parameters only poorly, since the specific profile is only relevant at very small scales (for a halo of mass $M = 10^{13}\,M_\odot h^{-1}$ at redshift $z = 2$, one finds that $1/r_s \simeq 25\,h\,\mathrm{Mpc}^{-1}$) and IM is mostly sensitive to larger scales \citep{Padmanabhan_etal_forecast}. 
Thus, these parameters are assumed to be fixed throughout this study.
The form of the concentration parameter assumes the same halo mass dependence for CDM as for neutral hydrogen. This choice is discussed in more detail in Section \ref{sec:axionsimprint}.

Ultimately, the Fourier transform of the radial density profile will be of importance
and is given by
\begin{align}
u_\hi(k|M) = \frac{4 \pi}{M_\hi(M)} \int_0^{R_v} \varrho_\hi(r,M) \frac{\sin kr}{kr} r^2 \diff r.
\label{eq:fourier_hi_profile}
\end{align}
For large $c_\hi$, it is well approximated by
\begin{align}
u_\hi(k|M) \simeq \frac{1}{(1+ (k r_s)^2)^2}.
\label{eq:uHI_approx}
\end{align}
All quantities in the present model are given in comoving coordinates (with exception of the virial velocity in equation \ref{eq:HI_halo_mass_relation}).
So, $r_s$ must be in comoving coordinates too and the physical virial radius in equation \eqref{eq:virial_radius} has to be converted, which, in effect, cancels the redshift dependence in the virial radius.

With this at hand, the Cartesian power spectrum is expressed by a one-halo term and two-halo term:
\begin{align}
P_\hi(k,z) = P_{\text{1h, \HI}} + P_{\text{2h, \HI}} 
\label{eq:halo_model_powerspectrum}
\end{align}
with
\begin{align}
P_{\mathrm{1h,\HI}}(k,z) = \frac{1}{\overline{\rho}^2_\hi} \int \diff M\,n(M,z) M_\hi^2(M,z) |u_\hi(k|M)|^2
\end{align}
and
\begin{align}
P_{\mathrm{2h,\HI}}(k,z) = b_\hi^2(k,z) P_{\mathrm{lin}}(k,z).
\label{eq:2-halo_term}
\end{align}
The \HI bias relates to the halo bias $b(M, z)$ and the halo mass function (HMF) $n(M,z)$ via
\begin{align}
b_\hi(k,z) = \frac{1}{\overline{\rho}_\hi} \int \diff M\,M_\hi(M,z) n(M,z) b(M,z) |u_\hi(k|M)|
\end{align}
and 
\begin{align}
\overline{\rho}_\hi = \int \diff M\,M_\hi(M,z) n(M,z).
\end{align}
Assuring $\overline{\rho}_\hi = \Omega_\hi \rho_{\mathrm{crit}}$ can be accomplished by fixing the normalization $\alpha$.
Note however that this normalization cancels for the above quantities and fixing $\overline{\rho}_\hi = \Omega_\hi \rho_{\mathrm{crit}}$ is computationally unnecessary.
For the calculation of the above quantities, we used the Sheth--Tormen HMF and bias \citep{Cooray_Sheth2002}. Both (HMF, $n$, and halo bias, $b$) are computed from the variance of matter fluctuations:
\begin{equation}
\sigma^2(R, z) = \frac{1}{2 \pi^2} \int_0^\infty P_{\mathrm{lin}}(k,z) W(k|R)^2 k^2 \diff k.
\label{eq:variance}
\end{equation}
For a spherical top-hat window function in real space, mass and radius are related by $M = 4 \pi \rho_0 R^3 / 3$ and the Fourier transform of that top-hat function is given by
\begin{align}
W(k|R) &= \frac{3}{(kR)^3} \left(\sin(kR) - kR \cos(kR)\right).
\label{eq:spherical_tophat_windowfunction}
\end{align}
The linear (matter) power spectra explicitly appearing in equations \eqref{eq:2-halo_term} and \eqref{eq:variance} were obtained from {\textsc{axioncamb}}.

\subsubsection{Axions and \HI}
\label{sec:axionsimprint}
In the previous sections, the modelling of the \SI{21}{cm} signal was discussed in rather general terms.
In this section we describe the way ULAs are accommodated in this formalism. 
In principle, to model their impact we have to consider its influence on the large-scale structure (i.e. specifically the HMF and halo bias), on small scales where they could influence the \HI density profile as well as onto the $M_{\hi}(M)$ relation:

\begin{enumerate} 
\item For the LSS, the main idea we employ is to treat ULAs below a certain mass similarly to massive neutrinos. 
For massive neutrinos, considerable effort has been put into investigating their influence on the HMF and bias: Large $N$-body simulations were employed and analysed in a series of papers \citep[e.g.][]{cosmo_neutrinosIII, cosmo_neutrinosII, cosmo_neutrinosI} and the spherical, top-hat collapse model was revisited for massive neutrinos in \citet{Ichiki_Takada2012}. 
Both studies found the halo mass function and halo bias are better fit if one considers only the baryon and CDM field, instead of the total matter field including neutrinos for their computation. 
Consequently, to model the \HI signal, one should take the \HI as a tracer of the CDM and baryon field, but not the total matter field including massive neutrinos. 
Note that neutrinos and axions are included in the dynamics of the perturbations and the background, and so affect the CDM + baryon fluid indirectly.

\citet{weighing_neutrinos2015} included massive neutrinos in modelling the \HI signal for low redshifts ($z<3$) in such a way and could show that constraints on the sum of the neutrino masses from late-time \SI{21}{cm} observations are possible.
To be precise, $\rho_0 = \rho_{\mathrm{crit}} ( \Omega_{\mathrm{CDM}} + \Omega_{b})$ was set throughout the calculation of the HMF and halo bias and for the computation of the variance (equation \ref{eq:variance}) and the two-halo term (equation \ref{eq:2-halo_term}), the power spectrum of the CDM and baryon component, $P_{\mathrm{CDM} + b}(k,z)$, was used. 
Furthermore, in the present case, $\rho_0$ was considered for the calculation of the virial radius (equation \ref{eq:virial_radius}). 
The choice on $\rho_0$ affects the \HI density profile (which is not significant as we will argue below) and the cut-off in the $M_\hi(M)$ relation. 
\begin{figure}
\centering
    \includegraphics[width=\linewidth]{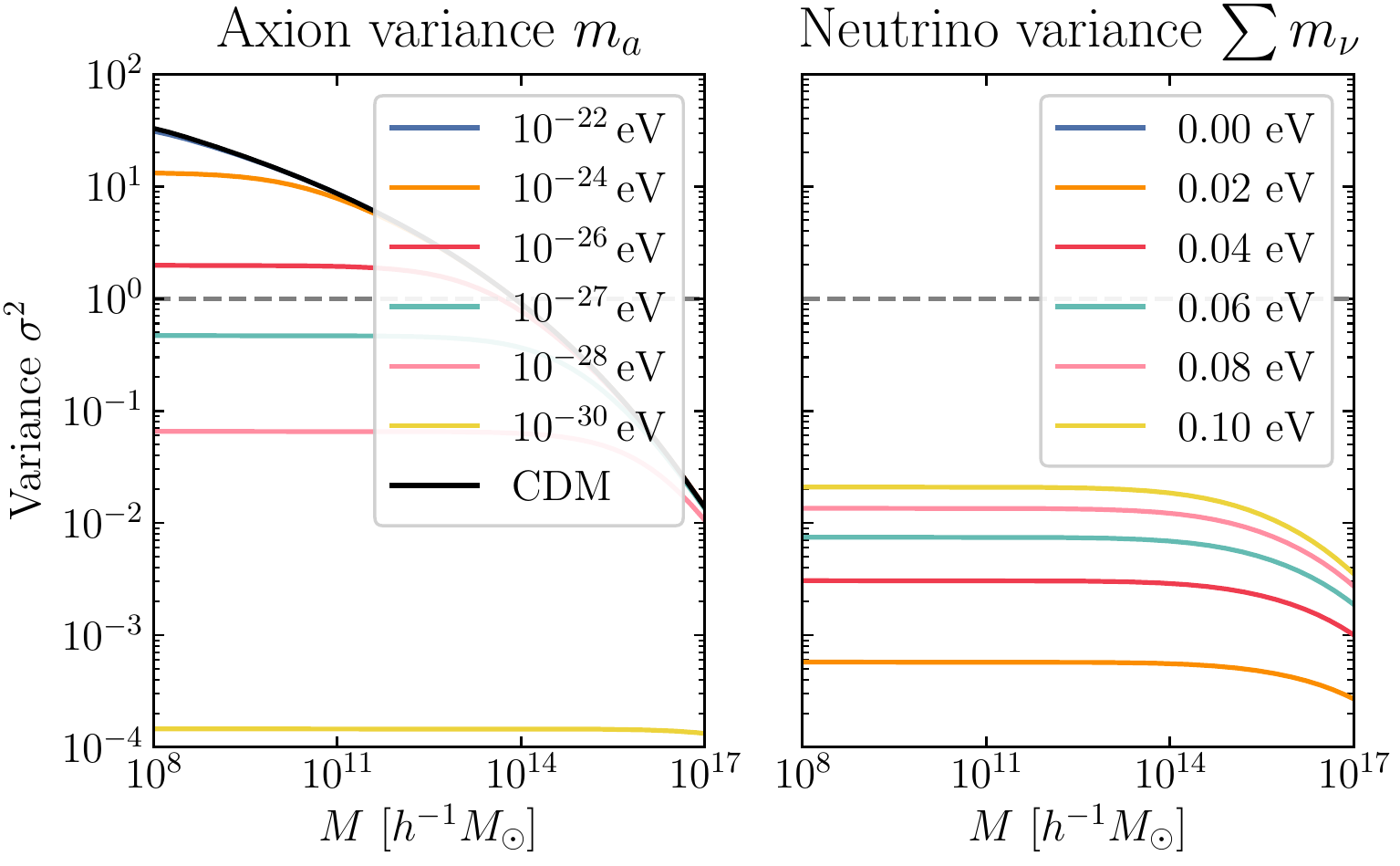}
    \caption{The variance with $P_{\mathrm{ax/}\nu}$ with parameters given in Table \ref{tab:parameters} and for the neutrino calculation for $\Omega_a/\Omega_d=0$. Halo masses are given with respect to the total matter background density. For axion masses $m_a < \SI{e-27}{eV}$ the variance is much smaller than 1, comparable to high neutrino masses and, as such, we treat such ULAs linearly, and do not include them into halos.}
    \label{fig:variance_plot}
\end{figure}

To relate axions of mass $m_a$ to massive neutrinos, we choose to compare the axion field variance to the neutrino field variance (see Fig. \ref{fig:variance_plot}):
For massive neutrinos, this is much lower than unity and, therefore, does not contribute significantly to the collapse of a halo. 
This justifies the linear approximation for neutrinos as not bound within halos, as discussed. 
Based on this observation, we choose to approximate ULAs in the same way as massive neutrinos whenever the variance is less than unity. 
In Fig. \ref{fig:variance_plot}, we observe that this is true for $m_a < \SI{e-27}{eV}$ at $z=0$.
For axions heavier than this boundary mass, $m_a \gtrsim \SI{e-27}{eV}$, we choose to treat them as usual CDM (i.e. collapsed into halos following the Sheth--Tormen model for the mass function), setting 
$\rho_0 = \rho_{\mathrm{crit}}( \Omega_{\mathrm{CDM}} + \Omega_{b} + \Omega_{a})$ and $P(k,z) = P_{b+\mathrm{CDM}+a}(k,z)$. 

\item The \HI--halo mass relation is subject to several {(astro-)physical} effects and absorbs those with a deterministic function of a few parameters.
It can be constrained empirically (fairly immune to the HMF and cosmology) and -- although not specifically including the possible impact of axions to it -- allowing it to vary with respect to its parameters.
A subtle issue is the choice of $\rho_0$ mentioned above, which affects the cut-off of \HI in equation \eqref{eq:HI_halo_mass_relation}.
To make the background density (and redshift) dependence on that cut-off clear, the exponential in equation \eqref{eq:HI_halo_mass_relation} can be rewritten in terms of the halo mass $M$ as $\exp\left[-M_{\mathrm{min}}(\rho_0, z)/M\right]$ with
\begin{equation}
M_{\mathrm{min}} = 2.45\times10^{10}\,M_\odot\,h^{-1} \left(\frac{v_{c,0}}{\SI{36.3}{km/s}}\right)^3 \left(\frac{0.3}{\Omega_0}\right)^{1/2} (1 + z)^{-3/2}.
\end{equation}
In effect, taking only the CDM+baryon component increases the cut-off mass when translating the exponential in the $M_\hi(M)$ relation (equation \ref{eq:HI_halo_mass_relation}) to halo masses compared to the case where $\rho_0$ includes the axions in addition. This is so, because of the fixed lower virial velocity $v_{c,0}$. 
Note, however, that this shift in the cut-off mass is small (for small axion and neutrino contributions) and a minor effect compared to the general impact of axions onto the \HI halo model and the overall uncertainty of $v_{c,0}$.
The fact that we marginalize over it in the Fisher forecast analysis should account for potential modelling uncertainties this choice introduces.

\item The \HI profile is affected by axions in its specific shape \citep[e.g. axions will alter the \HI profile on scales of order the de Broglie wavelength, where the ULA condenses into a soliton][]{Veltmaat_etal2019}.
Furthermore, the concentration parameter of \HI is assumed to be identical to that for CDM and, thus, the specific value of it also might depend weakly on axions. We note, however, that this form has a universal applicability in the description of low-$z$ surface density profiles and high-$z$ DLA observations~\citep{Padmanabhan_etal_forecast}. 
Generally, the \HI profile becomes relevant only at small scales (cf. Section \ref{sec:general_model}).
Since typically \HI IM surveys are concerned about larger scales, they are mostly insensitive to the specific \HI density profile and the instrumental noise is typically larger than the signal on those scales. 
Therefore, we neglect the specific impact of axions onto the \HI profile.
\end{enumerate}

Generally, we expect that the \HI halo model becomes less accurate on smaller, non-linear scales.
To account for this potential shortcoming, it is expedient to compare the results with a computation where the wavenumbers above a cut-off scale, $k_{\mathrm{nl}}$, are not considered. We adopt the redshift scaling as in \citet{Bull_Ferreira2015}: 
\begin{align}
k_{\mathrm{nl}} = \SI{0.14}{Mpc^{-1}} (1+z)^{2/(2 + n_s)},
\label{eq:k_nl}
\end{align}
where $n_s$ is the scalar spectral index.

\subsection{Results on the \HI Power Spectrum for ULAs}
\label{sec:results_HI_PS}
\begin{figure}
\centering
    \includegraphics[width=\linewidth]{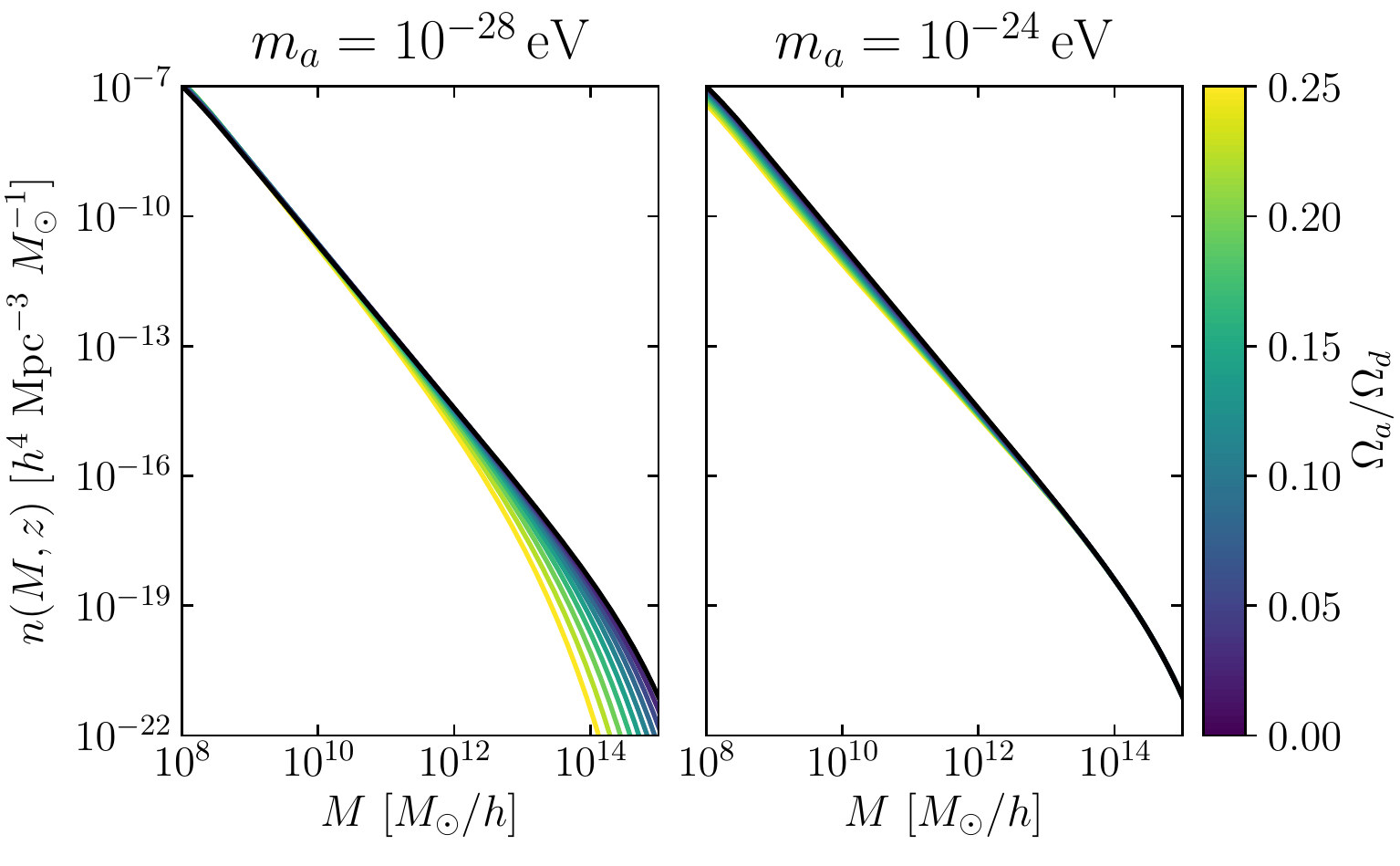}
    \caption{Sheth--Tormen halo mass function for $m_a = 10^{-28}$ and $\SI{e-24}{eV}$ and different axion fractions at $z=0$.}
    \label{fig:hmf_plot}
\end{figure}
\begin{figure}
\centering
    \includegraphics[width=\linewidth]{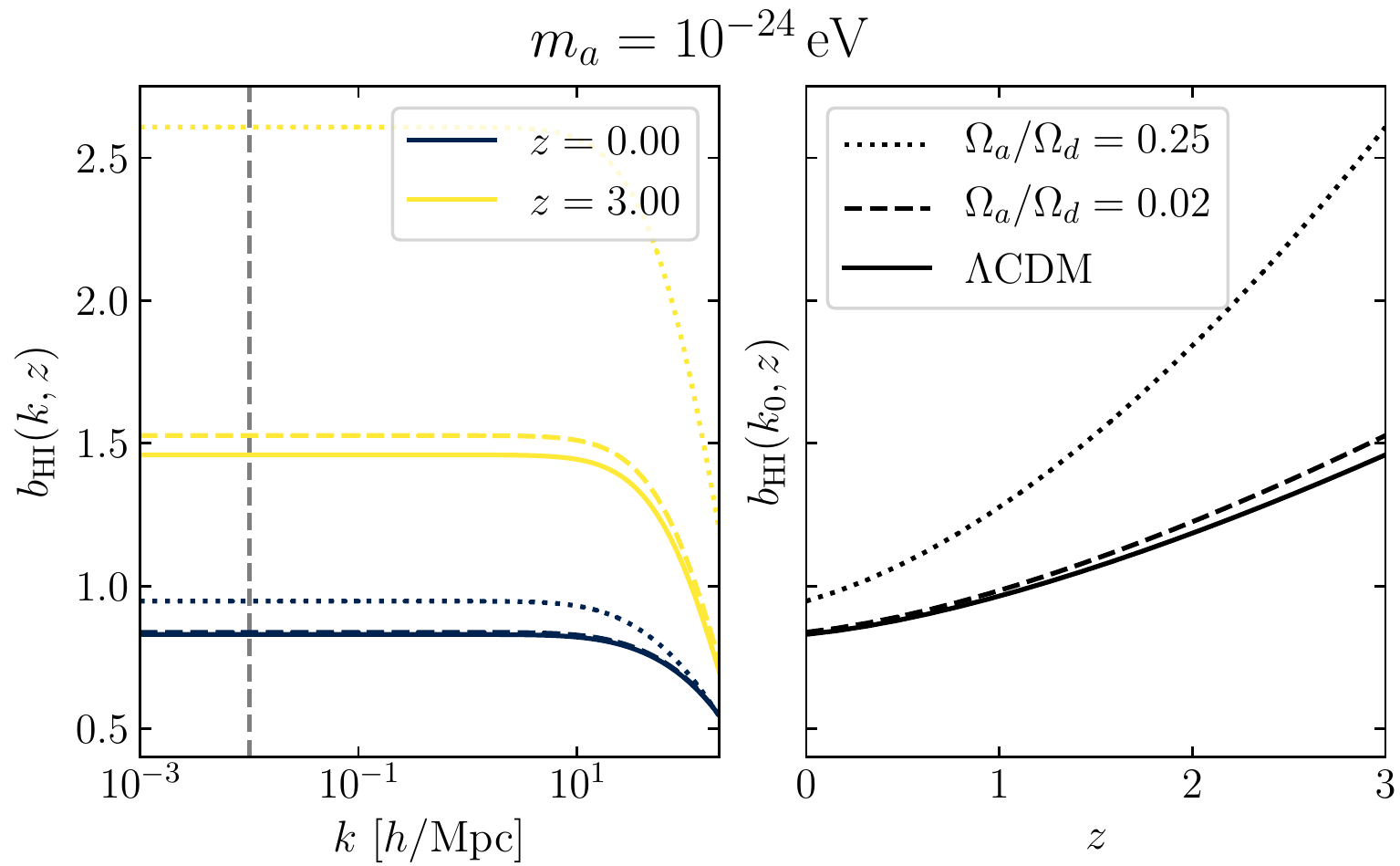}
    \caption{The left-hand panel shows the \HI bias as a function of comoving wavenumber $k$ at redshifts $0$ and $3$. The dashed, vertical line indicates $k_0 = 10^{-2}\,h{\mathrm{Mpc}}^{-1}$. The right-hand panel shows the \HI bias as a function of $z$ at $k_0$. Dashed (dotted) lines refer to a cosmology with a ULA of $m_a = \SI{e-24}{eV}$ and $\Omega_a/\Omega_d = 0.02$ ($\Omega_a/\Omega_d = 0.25$), whereas solid lines indicate the $\Lambda$CDM case.}
    \label{fig:b_HI}
\end{figure} 
\begin{figure*}
\centering
    \includegraphics[width=1.\linewidth]{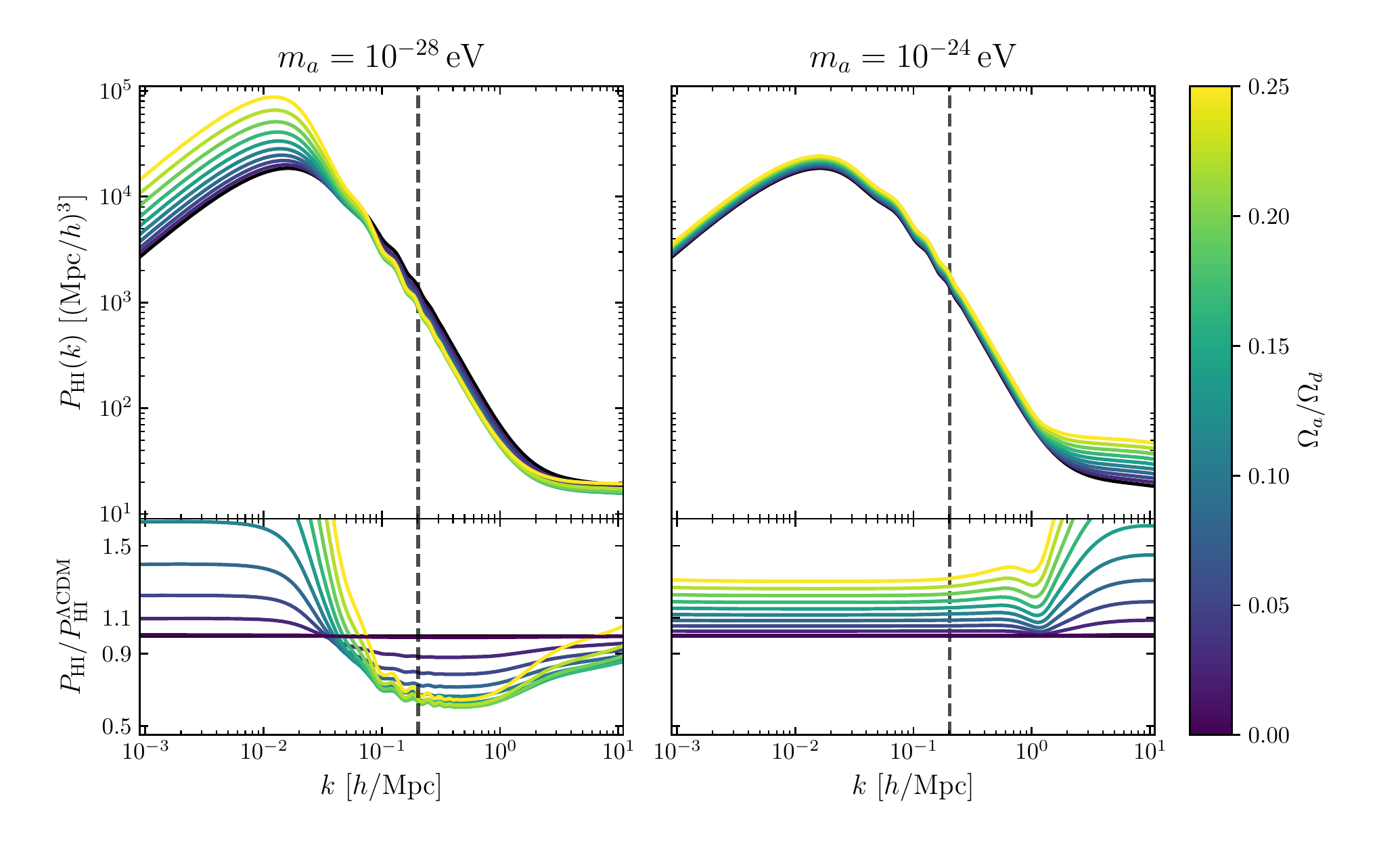}
    \caption{The 21\,cm power spectrum (equation \ref{eq:halo_model_powerspectrum}) for $m_a = 10^{-28}$ and $\SI{e-24}{eV}$ and different axion fractions at redshift 0. For higher axion masses, an overall enhancement is seen, whereas for lower axion masses power is suppressed at small scales due to the larger de Broglie wavelength (cf. \ref{sec:axion_physics}). The enhancement can be understood from the \HI bias in Fig. \ref{fig:b_HI}. The dotted line shows the non-linear scale ($k_{\mathrm{nl}} = \SI{0.14}{Mpc^{-1}}$ as defined in equation \ref{eq:k_nl}). The relative difference between the $\Lambda$CDM scenario and axion fractions at the $2\,\sigma$ exclusion limits obtained by the SKA1MID+CMB-SO surveys is on the per cent level.}
    \label{fig:P_HI}
\end{figure*} 
The HMF, the \HI bias, and the \HI power spectrum calculated with the current model are shown in  \Cref{fig:hmf_plot,fig:b_HI,fig:P_HI}, respectively. 
We restrict ourselves to two exemplary axion masses, each exhibiting the typical imprint on the \HI power spectrum in a certain range: 
first, a heavier axion, $m_a = \SI{e-24}{eV}$, treated as a CDM-like component, which is in its phenomenology similar to the fuzzy DM benchmark ($m_a \sim \SI{e-22}{eV}$), and, secondly, a lighter axion $m_a = \SI{e-28}{eV}$, which is in its impact similar to massive neutrinos (cf. \Cref{sec:axion_physics,sec:axionsimprint}) and, thus, is treated as a massive neutrino-like component.

Heavier ULAs, $m_a \sim \SI{e-24}{eV}$, suppress the formation of halos below their Jeans mass as shown in Fig. \ref{fig:hmf_plot}.
Due to the reduced number of low-mass halos, assuming a fixed \HI density parameter, the \HI has to reside in more massive halos that are more strongly biased.
Thus, the suppression of low-mass halos effectively leads to an enhancement in the \HI bias, which increases at higher redshift (cf. Fig. \ref{fig:b_HI}). 
This explains the main impact of axions in that mass range onto the \HI power spectrum (Fig. \ref{fig:P_HI}):
The suppression of the matter power spectrum is present only at small, non-linear scales for ULAs in that mass range.
Hence, when considering the two-halo term (equation \ref{eq:2-halo_term}) dominant on large scales, the main impact is the boost in the \HI bias.
This makes their imprint possibly degenerate with the astrophysical parameters controlling the \HI bias and $\Omega_\hi$ in general, which we discuss in more detail in \Cref{sec:degeneracy_structure,sec:discussion}.

In contrast, when $m_a = \SI{e-28}{eV}$ an enhancement of power is seen on large scales but a suppression on small scales is present compared to the $\Lambda$CDM case (cf. Fig. \ref{fig:P_HI}). 
These low-mass ULAs act dark energy (DE)-like, in the sense that they suppress matter fluctuations on almost all relevant scales. 
This shifts the HMF towards lighter halo masses, which leads to a significant decrease of intermediate halo masses ($\gtrsim 10^{11}\,M_\odot h^{-1}$ at $z=0.5$). 
Thus, more \HI needs to reside in each halo (similar to the effect of heavier ULAs) and the halo bias is also increased, leading to an enhancement in the \HI bias. 
This increase competes with the overall suppression of the linear power spectrum, which appears on already linear scales and reaches its saturation at the Jeans scale $k_J$ (cf. Section \ref{sec:axion_physics}). This gives rise to the scale-dependent imprint of lighter ULAs.

To summarize, we find that axions of all masses increase the \HI bias, albeit lighter ones out of a qualitatively different reason than heavier, fuzzy DM benchmark axions.
For lighter axions ($m_a \lesssim \SI{e-25}{eV}$), the suppression of the matter power spectrum competes with the boost in the \HI bias on already linear scales for the considered redshift range (i.e. 0--3 for realistic surveys), while for heavier ones, the suppression is ``hidden'' by the dominant one-halo term. 
Physically, this distinct behavior can be qualitatively understood by their different de Broglie wavelength. 
Lighter axions have a larger de Broglie wavelength and, therefore, smooth out matter power fluctuations on larger scales, ultimately given by $k_m$ and $k_J$ (cf. Fig. \ref{fig:scales_overview}).

Lastly, we compare our present findings with \citet{Carucci_2015}, who investigated the impact of warm dark matter (WDM) on the \SI{21}{cm} power spectrum with the help of $N$-body simulations \citep[similar to][]{Seehars_2016}. 
A subsequent paper with similar methodology from \citet{Carucci_2017} considers the impact of ULAs, when they are the only component in the DM sector.
Although we consider a mixed DM sector in this work, their findings are useful as they give guidance to what one should expect for high ULA density parameters for the fuzzy DM benchmark.
Reassuringly similarly an overall increase of the 21\,cm power spectrum for these heavier ULAs is found by \citet{Carucci_2017}
due to the fact that the formation of low-mass halos is suppressed and the \HI has to reside in the more massive halos that are more strongly biased. 

\begin{figure}
\centering
    \includegraphics[width=\linewidth]{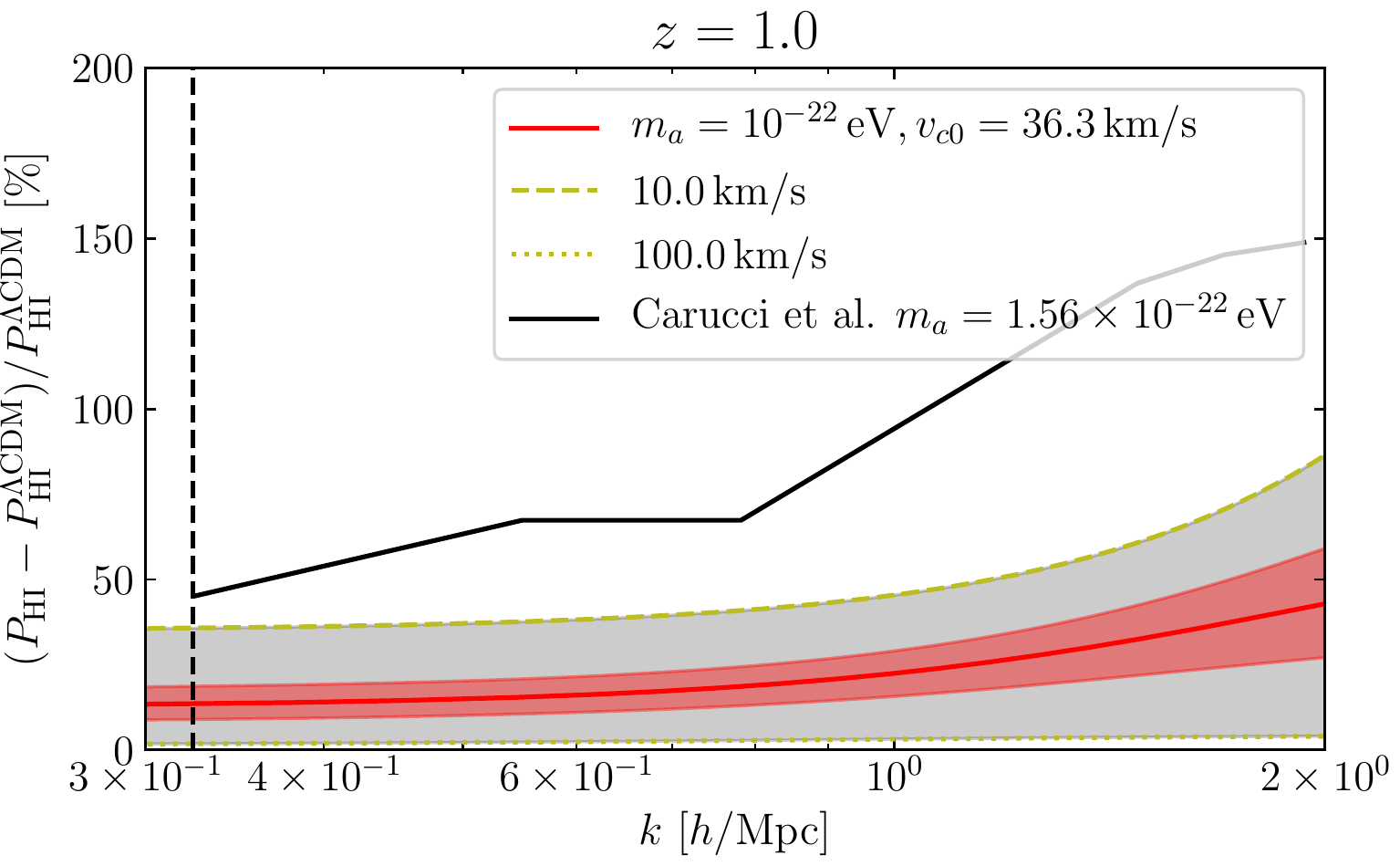}
    \caption{Relative difference in the \HI ($\SI{21}{cm}$) power spectrum given in per cent between $\Lambda$CDM and pure axion DM with $m_a = \SI{e-22}{eV}$ upon varying $v_{c,0}$. The red shaded region marks the $\pm 3\sigma$ confidence region around the best-fitting $v_{c,0}$ found in \citet{Refregier_Padmanabhan_Amara2017}. Other cosmological parameters are used as in Table \ref{tab:parameters}. The dotted vertical line refers to the non-linear cut-off wavenumber defined in equation \eqref{eq:k_nl}.}
    \label{fig:quant_comp_carucci}
\end{figure}

Fig. \ref{fig:quant_comp_carucci} shows the relative difference between $\Lambda$CDM and ULA only DM models at redshifts $z=1$ and compares to \citet[Fig. 7]{Carucci_2017}. 
While we observe a similar trend along $k$ and axion mass (not shown), our model underestimates the relative difference by a factor of a few relative to \citet{Carucci_2017}. 
Note that the $k$ range is already in the non-linear regime as given by equation \eqref{eq:k_nl}.
The increase of the relative difference for larger $k$ in our model comes about because of the one-halo term, which is dominant at those scales. 
Since for this term $M_\hi(M)$ goes in squared, the effect of ULAs is more pronounced.
It is reassuring that the one-halo and two-halo term seem to capture the trend observed by the simulation even in the mild non-linear regime.
On the other hand, we underestimate the relative difference between $\Lambda$CDM by a factor of a few.
Differences from our model are a different \HI--halo mass relation $M_\hi(M)$ \citep[which is redshift independent in the case of][]{Carucci_2017}, slightly different cosmological parameters and the inclusion of RSDs. 
Fig. \ref{fig:quant_comp_carucci} shows how the relative difference varies in our model for different $v_{c,0}$. 
It increases upon decreasing $v_{c,0}$ such that the suppression of low-mass halos becomes more important and might well explain some discrepancies between \citet{Carucci_2017} and our model. 
In short, the results are broadly consistent, given the overall methodological differences and that the simulations may have ingredients that are not necessarily tuned to match all the relevant data.
Since our model underestimates the impact of ULAs compared to that of \citet{Carucci_2017}, our forecasted constraints on the parameters are conservative.

\subsection{IM Observations}
\label{sec:experimental_setup}
In recent years several radio telescopes have been designed, planned (and constructed) to conduct 21\,cm IM surveys. 
In this work we specifically consider the future Square Kilometer Array \citep[SKA;][]{SKA_dewdney2013} telescope, its already built precursor, the MeerKAT telescope \citep{MeerKAT}, as well as the HIRAX \citep{HIRAX} and BINGO \citep{Battye_2012} telescopes. 
The important survey specifications for this work are listed in Table \ref{tab:experimental_setups}.
\begin{table*}
\caption{Instrumental parameters for different surveys from \citet{Bull_Ferreira2015} and for HIRAX from \citet{Ferreira2017}. The numbers below in parenthesis in column 6 and 7 (for $\nu_{\mathrm{max}}$ and $\nu_{\mathrm{min}}$) are the corresponding redshifts. We have combined band 1 and 2 of SKA1-MID as in \citet{Padmanabhan_etal_forecast}. The total observation time is set to $t_{\mathrm{tot}} = \SI{10000}{h}$ for all surveys.}
\begin{tabular}{l c c c c c c c c c}
\hline
\hline
Experiment &  {\makecell{$T_{\mathrm{inst}}$ \\ $[\mathrm{K}]$}} & {\makecell{$D_d$ \\ $[\mathrm{m}]$}} & {\makecell{$D_{\mathrm{min}}$ \\ $[\mathrm{m}]$}} & {\makecell{$D_{\mathrm{max}}$ \\ $[\mathrm{m}]$}} & {\makecell{$N_b$ \\ (dual \\ pol.)}} & $N_d$ &  {\makecell{$\nu_{\mathrm{max}}$ \\ $[\mathrm{MHz}]$}} & {\makecell{$\nu_{\mathrm{min}}$ \\ $[\mathrm{MHz}]$}} & {\makecell{$\Omega_{\mathrm{surv}}$ \\ $[\mathrm{sq.~deg.}]$}}\\ 
\hline
BINGO & $50$ & $25$ & - & - & $50 \times 2$ & $1$ & \makecell{$1260$ \\ $(0.13)$} & \makecell{$960$ \\ $(0.48)$} & $2000$ \\
MeerKAT (B1) & $29$ &  $13.5$ & - & -  & $1 \times 2$ & $64$ &  \makecell{$1015$ \\ $(0.4)$} & \makecell{$580$ \\ $(1.45)$} & $25000$ \\
SKA1-MID (B1+B2) & $28$ &  $15$ & - & -  & $1 \times 2$ & $190$ &  \makecell{$1420$ \\ $(0)$} & \makecell{$350$ \\ $(3.1)$} & $20000$ \\ 
HIRAX & $50$ &  $6$ & $6$ & $300$ & $1 \times 2$ & $1024$ &  \makecell{$800$ \\ $(0.78)$} & \makecell{$400$ \\ $(2.55)$} & $15000$\\
\hline
\hline
\end{tabular}
\label{tab:experimental_setups}
\end{table*}

A significant advantage of radio antennas compared to receivers in the optical range is that they can measure the phase of the incoming electromagnetic (EM) wave. Modern radio telescopes make heavily use of that and, therefore, typically consist of multiple dishes. Broadly, these can be run in two different modes: 
\begin{itemize}
\item \textit{Single dish:} One can autocorrelate the signal for each individual dish. 
This, effectively, increases the observation time for each pixel by the number of dishes (and number of beams). 
\item \textit{Interferometer:} 
The signal for each antenna is cross-correlated with another antenna, separated by a given baseline $\bmath{d}$. 
This, in result, increases the effective dish size for the antennas run in interferometric mode by the baseline, such that a much higher angular resolution is obtained. 
\end{itemize}

The radio telescopes are subject to thermal noise depending on the mode in which they are run. The equations used for the noise power spectrum of both modes are summarized shortly hereafter. 
Important specifications to calculate the noise of a radio telescope are the dish diameter of a single dish $D_d$, the number of dishes $N_d$, the number of beams $N_b$ (which includes the number of different polarization channels $n_{\mathrm{pol}}$, which generally equals two), the frequency channel width (here corresponding to the redshift bin width) $\Delta \nu$, the solid angle sky coverage of the survey $\Omega_{\mathrm{surv}}$, and the observed wavelength of the incoming EM wave $\lambda$. 
Furthermore, the total system temperature is taken $T_{\text{sys}} = T_{\text{sky}} + T_{\text{inst}}$, with $T_{\mathrm{sky}} = \SI{60}{K} \left(\SI{350}{MHz}/\nu\right)^{2.5}$ \citep{Padmanabhan_etal_forecast}.
\subsubsection{Single-dish mode}
For the single-dish mode, we used the noise expression given in \citet{Knox1995}.
Together with the beam smearing term, the dimensionless noise power spectrum is given by 
\begin{equation}
N_\ell = \frac{1}{\bar{T}_b^2} \Omega_{\text{pix}} \sigma_{\text{pix}}^2 \exp\left[\frac{\ell^2 \theta_B^2}{8 \ln{2}}\right],
\label{eq:Knox_noise}
\end{equation}
where $\theta_B$ is the beam full width at half-maximum $\theta_B \approx \lambda/D_d$, $\Omega_{\text{pix}}$ is the solid angle beam area, and $\bar{T}_b$ is the mean brightness temperature of expected \HI signal given by \citep{Villaescusa-Navarro_etal_2016}
\begin{align}
\bar{T}_b \simeq 190\,\Omega_\hi h \,\frac{(1 + z)^2}{E(z)}\,\si{mK}.
\label{eq:mean_brightness_temperature}
\end{align}
The thermal noise per pixel is given by the radiometer equation. For a perfect receiver, one has \citep{Battye_2012}:
\begin{align}
\sigma_{\text{pix}} = \frac{T_{\text{sys}}}{\sqrt{t_{\text{pix}} \Delta \nu}}.
\label{eq:sigma_pixel}
\end{align}
The time of observation for each pixel depends on the total observation time divided by the number of pixels in the map and multiplied by the number of dishes and beams (if the radio telescope is run in an autocorrelation mode). Therefore, it is
\begin{equation}
t_{\text{pix}} = t_{\text{obs}} N_d N_b \frac{\Omega_{\text{pix}}}{\Omega_{\text{surv}}}.
\label{eq:pixel_time}
\end{equation}

Note that this noise expression equals the noise expression given in \citet{Bull_Ferreira2015} upon converting into angular scales except of a different prefactor $\sim 3$ due to the inclusion of the effective dish area (cf. Appendix \ref{sec:appendix} for more details on the conversion and their similarity).
\subsubsection{Interferometric Mode}
\label{sec:interferometer}
The interferometric noise expression is subject to several technical uncertainties and different expressions are used in the literature for upcoming radio telescopes. 
Specifically, \citet{Jalilvand_etal_2020} discusses the thermal noise properties for the HIRAX survey, where an ``optimistic''  expression was compared to and ``pessimistic'' one. Furthermore, an expression is included which they note as ``realistic'' since it matches recent simulations for the HIRAX interferometer noise. 
According to \citet{Jalilvand_etal_2020} the difference between the optimistic and pessimistic scenarios arises due to the assumption of how the survey area is scanned: at once (instantaneous) or sequentially. 
We choose the interferometric noise expression to be identical to the optimistic case of \citet{Jalilvand_etal_2020}, since the S/N seems to align more for the optimistic and realistic scenarios.\footnote{Note that, in the code published along with this paper, both ``optimistic'' and ``pessimistic'' modes are implemented.}
It is given by \citep{Pourtsidou2016}
\begin{align}
N_{\ell}^{\mathrm{interf.}} = \frac{T_{\mathrm{sys}}^2 {\mathrm{FOV}}^2 }{\bar{T}_b^2 n_{\mathrm{pol}} \Delta \nu t_{\mathrm{tot}} n(\bmath{u} = \ell / 2 \pi)},
\label{eq:noise_interferometer}
\end{align}
with the field of view, ${\mathrm{FOV}} \approx \lambda^2/D_d^2$, and the baseline distribution in the $uv$-plane $n(\bmath{u})$ is taken from \citet{Bull_Ferreira2015} if available for the specific survey. 
If no baseline distribution is available the following approximate formula is used \citep{Bull_Ferreira2015}:
\begin{align}
n(\bmath{u}) = \frac{N_d (N_d - 1)}{2 \pi (u_{\mathrm{max}}^2 - u_{\mathrm{min}}^2)},
\end{align}
where $u_{\mathrm{max/min}} = D_{\mathrm{max/min}} / \lambda$ and $D_{\mathrm{max/min}}$ is the maximum/minimum diameter of the array. For $u = \ell/(2 \pi)$ above or below $u_{\mathrm{max/min}}$ the noise is set to infinity. 

\begin{figure}
    \includegraphics[width=\linewidth]{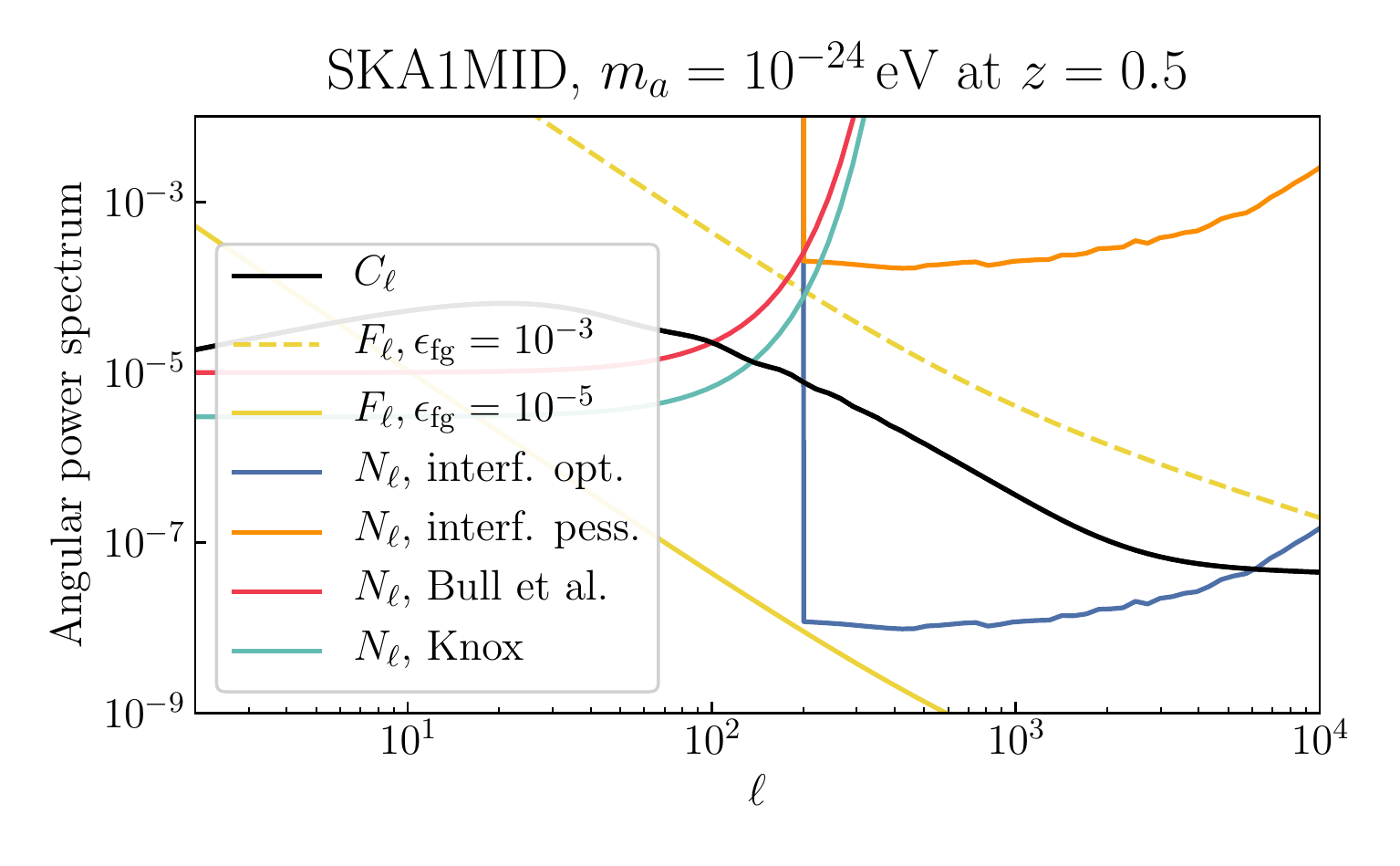}
    \caption{Dimensionless angular power spectrum of the 21\,cm signal, the residual foreground and of the noise with different expressions. For the interferometric mode both pessimistic and optimistic cases for the noise are shown, while the optimistic expression was used throughout this study as discussed in Section \ref{sec:interferometer}. The signal power spectrum is calculated for $m_a = \SI{e-24}{eV}$ and fiducial parameters as in Table \ref{tab:parameters}.}
    \label{fig:noise_summary}
\end{figure}

The overall dimensionless angular noise power spectra are shown in Fig. \ref{fig:noise_summary}, together with the 21\,cm angular power spectrum for the SKA1MID configuration at redshift $z=0.5$. 
\subsubsection{Foregrounds}
\label{sec:foregrounds}
It is expected that foregrounds are the leading systematic effect for \HI IM surveys. 
However, it is possible to remove them from the overall signal because of their spectrum, i.e. foregrounds depend differently on frequency than the \HI signal. 
This removal procedure effectively renders some modes in the radial direction, $k_\parallel$,  useless to the analysis, but ``cleans'' the signal from foregrounds. 

For our forecasts, foregrounds have been neglected and a perfect foreground removal has been assumed. 
However, to get an idea on their influence and importance, we considered the foreground components for the equal-$\nu$ power spectrum \citep{Santos_Cooray_Knoxs_2005} with an effective foreground residual amplitude $\epsilon_{\mathrm{fg}}$ (which is related to the number of modes in radial direction for the foreground removal, $k_\parallel < k_{\parallel}^{\mathrm{fg}}$) as in \citet{Bull_Ferreira2015}: 
\begin{equation}
F_\ell = \epsilon_{\mathrm{fg}}^2 \sum_X A_X \left(\ell_*/\ell \right)^{\beta_X} \left(\nu_*/\nu\right)^{2 \alpha_X}
\end{equation}
with $\ell_* = 1000$ and $\nu_* = \SI{130}{MHz}$ and parameters $A_X,~\alpha_X,$ and $\beta_X$ as in \citet[][table 1]{Santos_Cooray_Knoxs_2005}.

In Fig. \ref{fig:noise_summary}, the dimensionless residual power spectrum can be seen for the different components. 
For efficient foreground removal with $\epsilon_{\mathrm{fg}} \lesssim 10^{-5}$, the foreground contamination is subdominant to the thermal noise for large $\ell$. We conclude that in this case, the results of the forecasts would not change qualitatively. 
However, this conclusion changes drastically if less efficient foreground removal is assumed, when the foreground residuals dominate the cosmological signal and noise altogether. 
Note that these considerations did not include the removal of foreground-contaminated modes in the radial direction ($k_{\parallel} < k_{\parallel}^{\mathrm{fg}}$) or the ``foreground wedge'' \citep{Ferreira2017}. 
Therefore, Fig. \ref{fig:noise_summary} simply highlights the importance of effective foreground removal, requiring $\epsilon_{\mathrm{fg}} \lesssim 10^{-5}$ (from the requirement $F_\ell \lesssim N_\ell$).
\subsection{Fisher matrix forecasts}
\label{sec:fisher}
To assess the viability of future 21\,cm IM surveys to constrain the cosmological parameters in general and the fractional axion density parameter $\Omega_a/\Omega_d$ in particular we develop Fisher matrix forecasts.
The inverse Fisher matrix $\mathbfss{F}^{-1}$ is the covariance matrix of the probability distribution of the parameter and gives an estimate on the best possible, minimal error on the parameters \citep{Tegmark1997}.

Assuming a Gaussian likelihood with covariance matrix $C$, one can show that the Fisher information matrix takes the following form \citep{Tegmark1997}:
\begin{align}
F_{ij} = \frac{1}{2} \mathrm{tr}\left[\mathbfss{C}^{-1} \frac{\partial \mathbfss{C}}{\partial p_i} \mathbfss{C}^{-1} \frac{\partial \mathbfss{C}}{\partial p_j} \right] + \frac{\partial \bmath{\mu}^T}{\partial p_i}\mathbfss{C}^{-1}\frac{\partial \bmath{\mu}}{\partial p_j},
\end{align}
where $\bmath{p}$ denote the model parameters, and $\bmath{\mu}= \langle \bmath{x} \rangle$.
In the present case, $\bmath{x}$ are the temperature fluctuations expanded in spherical harmonics and the forecasted model parameters are $\bmath{p}~=~\{\ln{A_s}, n_s, \Omega_b, \Omega_c, \Omega_a/\Omega_d, h, v_{c,0}, \beta\}$.
We assume isotropy such that
\begin{align}
\mu &= 0 \\
C_{ij} &= \delta_{ij} \left[C_\ell + N_\ell\right].
\end{align}
With that at hand, the Fisher matrix is given by
\begin{align}
F_{ij} &= \sum_\ell \frac{1}{(\Delta C_\ell)^2} \pd{C_\ell}{p_i} \pd{C_\ell}{p_j},\\
(\Delta C_\ell)^2 &= \frac{2}{(2\ell +1)f_{\mathrm{sky}}}(C_\ell + N_\ell)^2,
\label{eq:final_fisher_matrix}
\end{align}
where the factor $f_{\mathrm{sky}}$ is included to account for the fact the survey is scanning only a fraction of the sky $f_{\mathrm{sky}} = \Omega_{\mathrm{surv}}/(4\pi)$. In the above expression, it was assumed that the noise $N_\ell$ does not depend on any parameters that shall be constrained/forecasted. 
This is not strictly true in the dimensionless framework as the mean brightness temperature depends weakly on cosmological parameters (cf. equation \ref{eq:mean_brightness_temperature}). We will, however, assume that the mean brightness temperature is fixed, e.g. determined from other surveys/probes and neglect its information to the Fisher matrix in the present analysis \citep[cf.][]{Padmanabhan_etal_forecast, Chen_Battye_forecast}.

\begin{figure*}
\includegraphics[width=.45\linewidth]{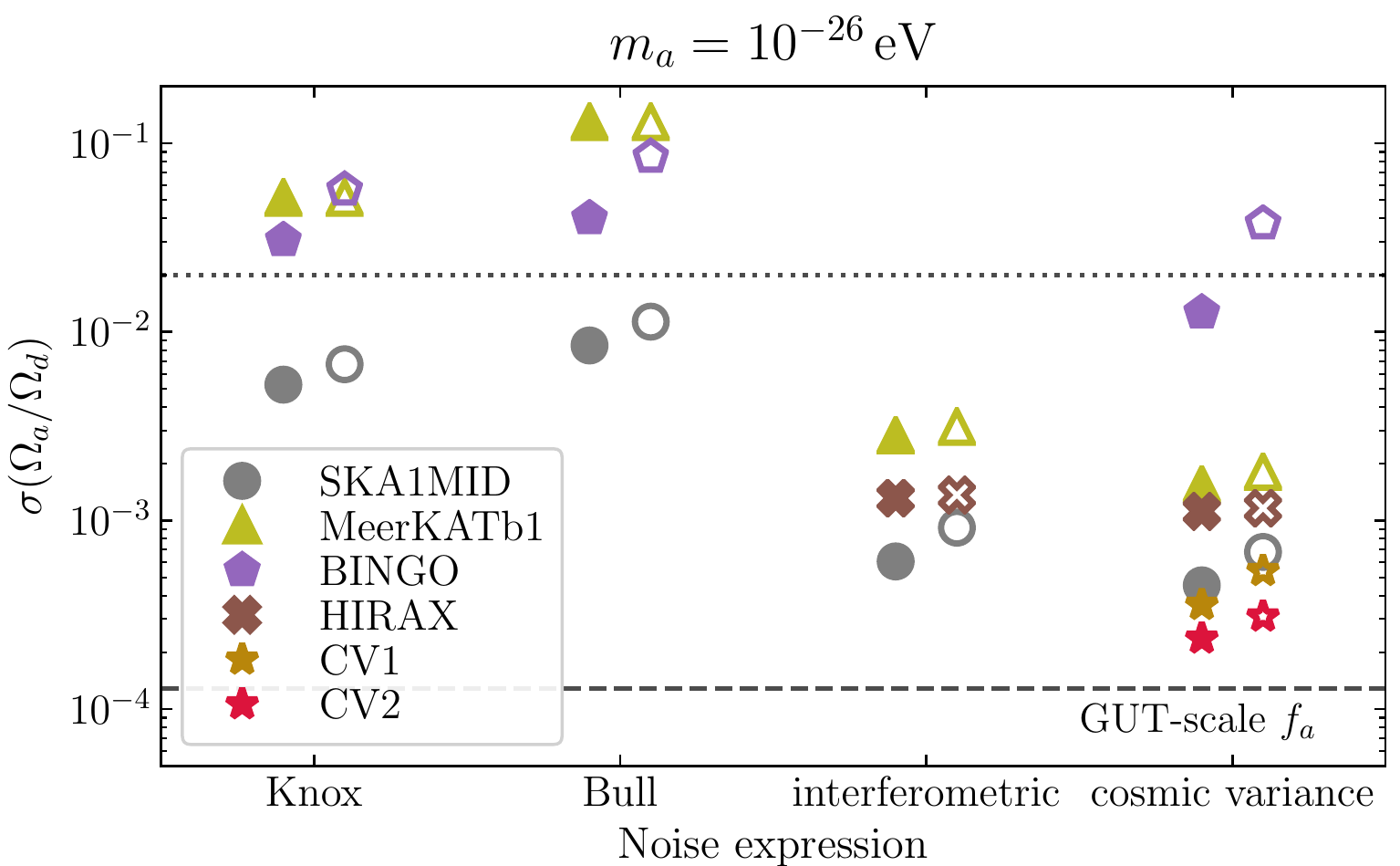} \hspace*{0.2cm} \includegraphics[width=.45\linewidth]{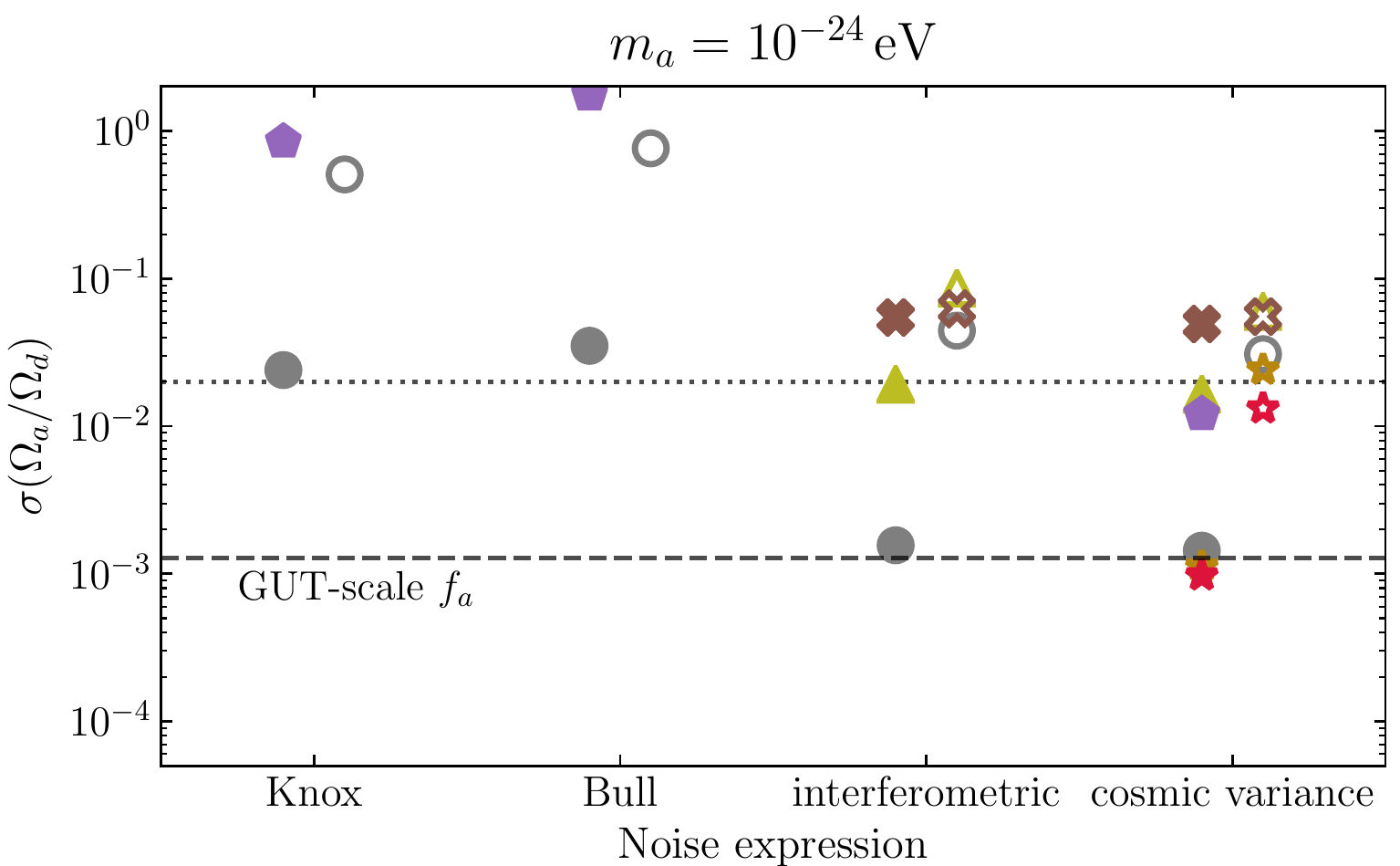}
    \caption{Scan over all surveys and noises including all scales (full markers) and when only linear scales are included (empty markers). ``cosmic variance'' denotes cosmic variance limited surveys by switching off instrumental noise. The different surveys are then only distinguished by the redshift range and sky coverage. On the $y$-axis the marginal error obtained from the cumulative Fisher matrix is shown. The dashed line indicates the axion fraction obtained at the GUT scale and the dotted line indicates the fiducial axion abundance. The left-hand panel considers an axion of mass $m_a = \SI{e-26}{eV}$ and the right-hand panel $m_a = \SI{e-24}{eV}$. CV1 and CV2 are mock surveys with $f_{\mathrm{sky}} = 1$ and for a redshift range of [0,3] and [0,5], respectively. For lighter axions $m_a \lesssim \SI{e-25}{eV}$ the scale-dependent imprint takes place on linear scales. Thus, degeneracies are already broken on those scales and the exclusion of non-linear scales does not alter the constraints significantly (left-hand panel). This stands in contrast to heavier axions $m_a \gtrsim \SI{e-24}{eV}$ (right-hand panel) where non-linear scales are crucial to break degeneracies with the astrophysical parameters and $A_s$.}
    \label{fig:survey_scan_linear}
\end{figure*}
 
As mentioned above, the redshift range will be divided into several bins. For each bin $i$, a Fisher matrix $\mathbfss{F}^{(i)}$ is obtained. 
In the Limber approximation, cross-correlations between different redshift bins have been effectively neglected. 
In this picture, one can then simply add the Fisher matrices from each redshift bin to obtain the cumulative Fisher matrix, containing all the information that can be gained from the survey in the current formalism:
\begin{align}
\mathbfss{F}_{\mathrm{cumul}} = \sum_{i = 1}^{N_{\mathrm{bin}}} \mathbfss{F}^{(i)}.
\end{align} 
We take equal-sized redshift bin width of $\Delta z = 0.05$ and calculate the $C_\ell$s in equation \eqref{eq:dimensionless_Cl_with_Limber} at the mid-points of each. Furthermore, we restrict ourselves to $\ell \leq 1000$ when forecasting.

The partial derivatives in equation \eqref{eq:final_fisher_matrix} have been calculated numerically by using the central finite difference 
\begin{align}
\frac{\partial f(x)}{\partial x} = \frac{f(x + \Delta x) - f(x - \Delta x)}{2 \Delta x}.
\end{align}
The step sizes for calculating the derivatives have been carefully chosen such that a sufficient level of convergence was reached. 
The fiducial values for the forecasted parameters, as well as the step size for calculating the derivatives, are listed in Table \ref{tab:parameters}. 
The general outline of the code written for this analysis is based on the code presented in \citet{Bull_Ferreira2015}.
\footnote{The adapted version for this analysis is publicly available at \url{https://github.com/JurekBauer/axion21cmIM.git}.}
We conclude by listing the main assumptions taken into account for the model presented in this section:
\begin{enumerate}
\item All angular power spectra are calculated using the Limber approximation.
\item No peculiar velocities, i.e. RSDs, are included in the calculation for the \HI angular power spectrum.
\item $\Omega_\hi$ is supposed to be fixed and redshift-independent.
\item Foregrounds can be removed efficiently and do not intercept with the \HI signal.
\item Axions can be modeled like massive neutrinos for $m_a < \SI{e-27}{eV}$ and as a CDM-like component for $m_a \gtrsim \SI{e-27}{eV}$.
\item No cosmological information gained from $\bar{T}_b$.
\item Cosmological information can be obtained from $b_\hi$ (the \HI bias is not treated as model-independent or a ``nuisance'' parameter).
\item Axions do not change the neutral hydrogen profile.
\item No further inclusion of nuisance parameters, which are marginalized over, than those given in above and listed in Table \ref{tab:parameters}.
\end{enumerate}

\section{Results}
\label{sec:results}
\subsection{Survey Comparison}
To compare different surveys, we look at how they constrain $m_a = 10^{-26}$ and $m_a = \SI{e-24}{eV}$ on the axion fraction. We choose the first mass since it is the most constrained bin in the CMB analyses \citep{Hlozek_Marsh2018} and the second to highlight the mass-dependent impact discussed in Section \ref{sec:results_HI_PS}.
The estimated marginal error on $\Omega_a/\Omega_d$ is shown for different configurations in Fig. \ref{fig:survey_scan_linear}.\footnote{Note that marginal errors larger than the fiducial value imply {$\Omega_a/\Omega_d<0$} within the Fisher forecast analysis, which is unphysical. Since we generally observed that the bounds do not change significantly with the fiducial axion fraction, we do not expect that the bounds are affected strongly by the inclusion of the physical prior $\Omega_a/\Omega_d>0$.}
CV1 and CV2 denote cosmic variance limited surveys (the noise is set to zero) for redshift ranges [0,3] and [0,5], respectively, and $f_{\mathrm{sky}} = 1$. 
Depending on the instrumental noise properties CV2 might be comparable to PUMA \citep{PUMA}.
We also consider the case of making each survey in turn ``cosmic variance limited'' by switching off instrumental noise. The different surveys are then only distinguished by the redshift range and sky coverage. 

The filled markers in Fig. \ref{fig:survey_scan_linear} include all scales accessible by the survey, while the empty ones use the non-linear cut-off scale in equation \eqref{eq:k_nl}:
The instrumental noise is set to infinity for $\ell > k_{\mathrm{nl}} r(z)$. 
In this way, the non-linear scales are excluded and do not pass any information to the Fisher matrix. 
Commonly, radio telescopes run in the interferometric mode are more constraining on the axion fraction and come close to cosmic variance limited surveys when non-linear scales are included (this, however, might change when allowing for $\ell > 1000$). 
Specifically, SKA1MID is much more effective if run in the interferometric mode (cosmic variance limited) than in the single-dish mode. This is even more marked for MeerKAT.
Generally and in realistic instrumental noise scenarios, HIRAX and SKA1MID run in the interferometric mode are the most constraining surveys.  

As shown in Fig. \ref{fig:noise_summary}, radio telescopes, run in an interferometric mode, typically scan smaller and, thus, potentially non-linear scales. 
Evidently, excluding non-linear scales leads to an increase for the error on the axion fraction for all experiments and both axion masses. 
However, the magnitude strongly depends on the axion mass: While constraints for $m_a = \SI{e-26}{eV}$ are barely affected, constraints for $m_a = \SI{e-24}{eV}$ increase significantly upon excluding non-linear scales. 
This can be understood by the fact that degeneracies with other parameters are already broken on linear scales for the lighter ULAs. 
Interestingly, the HIRAX survey is an exception to that observation, such that we conclude that the HIRAX survey is only mildly sensitive to non-linear scales.
To shed more light onto the degeneracy structure, it is discussed in the next section for heavy mass ULAs in more detail.
\subsection{Degeneracy Structure}
\label{sec:degeneracy_structure}
\begin{figure}
\centering
    \includegraphics[width=1.\linewidth]{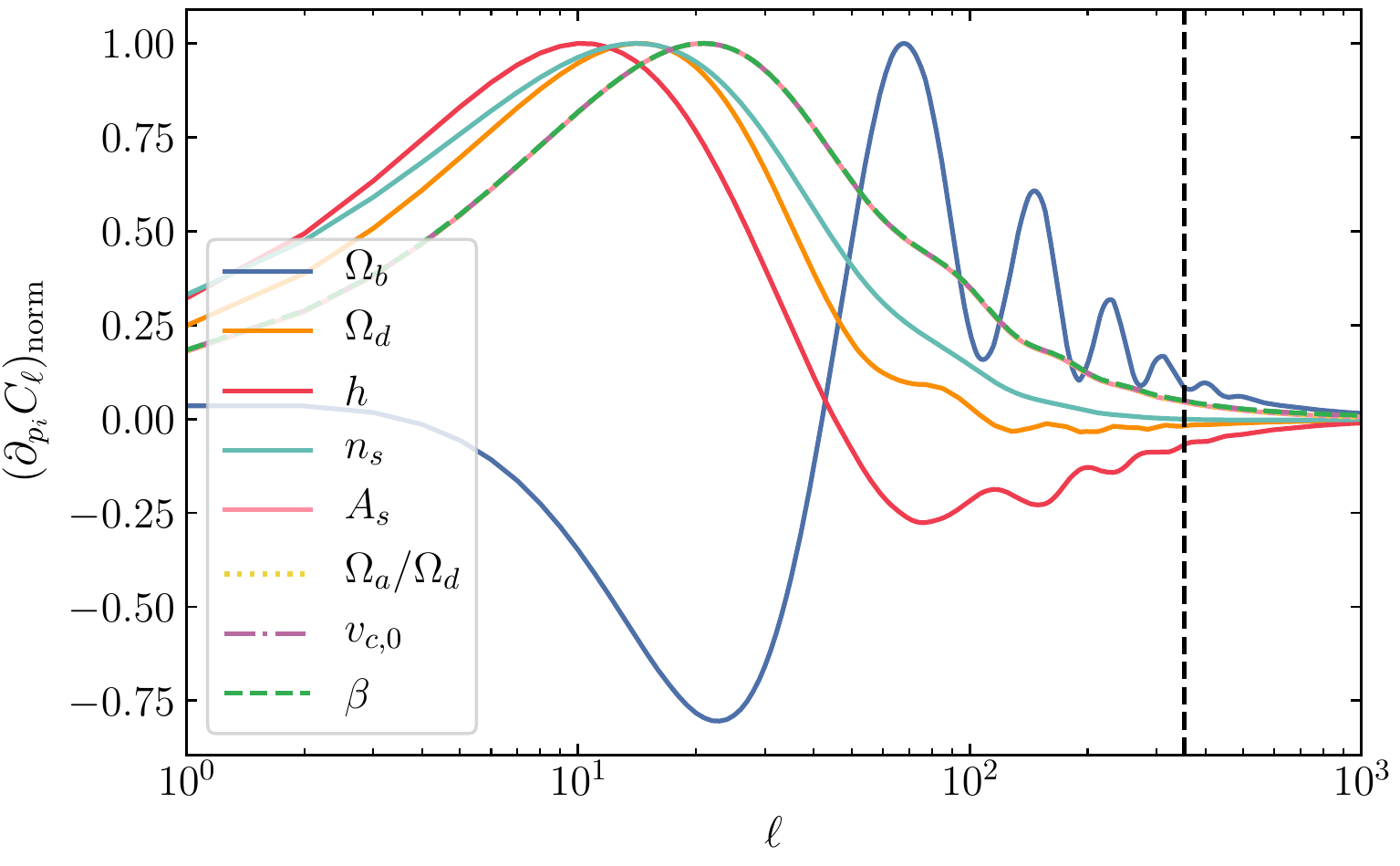}
    \caption{Normed Fisher derivatives (equation \ref{eq:normed_derivatives}) for $m_a = \SI{e-24}{eV}$ and $\Omega_a/\Omega_d = 0.02$ at $z=0.5$. The black, dashed line indicates $\ell$ corresponding to $k_{\mathrm{nl}}$ (equation  \ref{eq:k_nl}), the non-linear cut-off scale. Note that the curves $A_s, \Omega_a/\Omega_d, v_{c,0},$ and $\beta$ are all coincident in this figure.}
    \label{fig:derivative_comparison}
\end{figure}
To study the degeneracy structure of the \HI IM survey with ULAs, the derivatives $\partial_{p_i} C_\ell (:= C_\ell^\prime$ here) in the definition of the Fisher matrix (equation \ref{eq:final_fisher_matrix}) are of importance. 
Fig. \ref{fig:derivative_comparison} shows the normed derivatives defined by
\begin{align}
(C_\ell^\prime)_{\mathrm{norm}} = C_\ell^\prime \cdot \begin{cases} \max_{\ell}(C_\ell^\prime)^{-1} &  \text{if } |\max_{\ell}(C_\ell^\prime)| > |\min_{\ell}(C_\ell^\prime)| \\
\min_{\ell}(C_\ell^\prime)^{-1} & \text{else,}
\end{cases}
\label{eq:normed_derivatives}
\end{align}
for $m_a = \SI{e-24}{eV}$ and $\Omega_a/\Omega_d = 0.02$ at $z=0.5$. 
While the derivative with respect to parameters $\Omega_d,~\Omega_b,~h,$ and $n_s$ have distinct shapes, the four parameters $A_s,~\Omega_a/\Omega_d,~\beta,$ and $v_{c,0}$ resemble each other closely. 
These degeneracies are expected: 
The astrophysical parameters $\beta$ and $v_{c,0}$ dictate the specific value of $b_\hi$ (and for large scales less importantly the one-halo term). 
Similarly for $m_a \gtrsim \SI{e-24}{eV}$ and on the considered scales, $\Omega_a/\Omega_d$ engenders a scale-independent enhancement in the 21\,cm signal because of the boost in $b_\hi$ (cf. Fig. \ref{fig:b_HI} and the discussion in Section \ref{sec:results_HI_PS}). 
$A_s$ does alter $b_\hi$ and the matter power spectrum linearly, and thus an increase in $A_s$ also results in a scale-independent increase in the 21\,cm signal. 
Subsequently, all these parameters induce -- if one is only concerned about large, linear scales -- a scale-independent imprint and the normed derivatives are similar. 

There are several ways to break these degeneracies: (i) Providing a prior from other observations for the cosmological parameters (e.g. the CMB fluctuations for $A_s$ and $\Omega_a/\Omega_d$), (ii) including RSDs \citep[isolating $A_{s}$;][and possibly $\Omega_a/\Omega_d$, depending on the influence of axions onto RSDs]{Chen_Battye_forecast}, (iii) probing non-linear scales (i.e. the one-halo term) , (iv) the different redshift scaling of the impact of each parameter, or
(v) independent observations to constrain $v_{c,0}$ and $\beta$ tightly.
Probing non-linear scales breaks the degeneracy because, on smaller scales, the one-halo term becomes relevant which is affected differently for the parameters (e.g. the magnitude in increase is larger for $A_s$ and $\Omega_a/\Omega_d$ than for the astrophysical parameters) and depending on the axion mass and abundance the suppression at the Jeans scale in the two-halo term is visible.
Since RSDs are not included at present, degeneracies between those four parameters are (partly) broken by (i), (iii), (iv), and (v) in what follows. 

Note that Fig. \ref{fig:derivative_comparison} also shows that degeneracies with other parameters are likely, although less marked: For example, a degeneracy with $\Omega_b$ is present if one is only concerned about smaller scales and the BAOs are washed out.

The degeneracy of axions with CDM is relevant for other cosmological probes \citep[e.g. the CMB][]{Amendola_Barbieri2006, Hlozek_Marsh2015} and also has an interesting structure for \HI IM. 
For axions of high-mass $m_a \gtrsim \SI{e-24}{eV}$, a degeneracy with $\Omega_{\mathrm{CDM}}$ (if varied along with $\Omega_\Lambda$) is expected on large, linear scales. 
\begin{figure}
\centering
    \includegraphics[width=1.\linewidth]{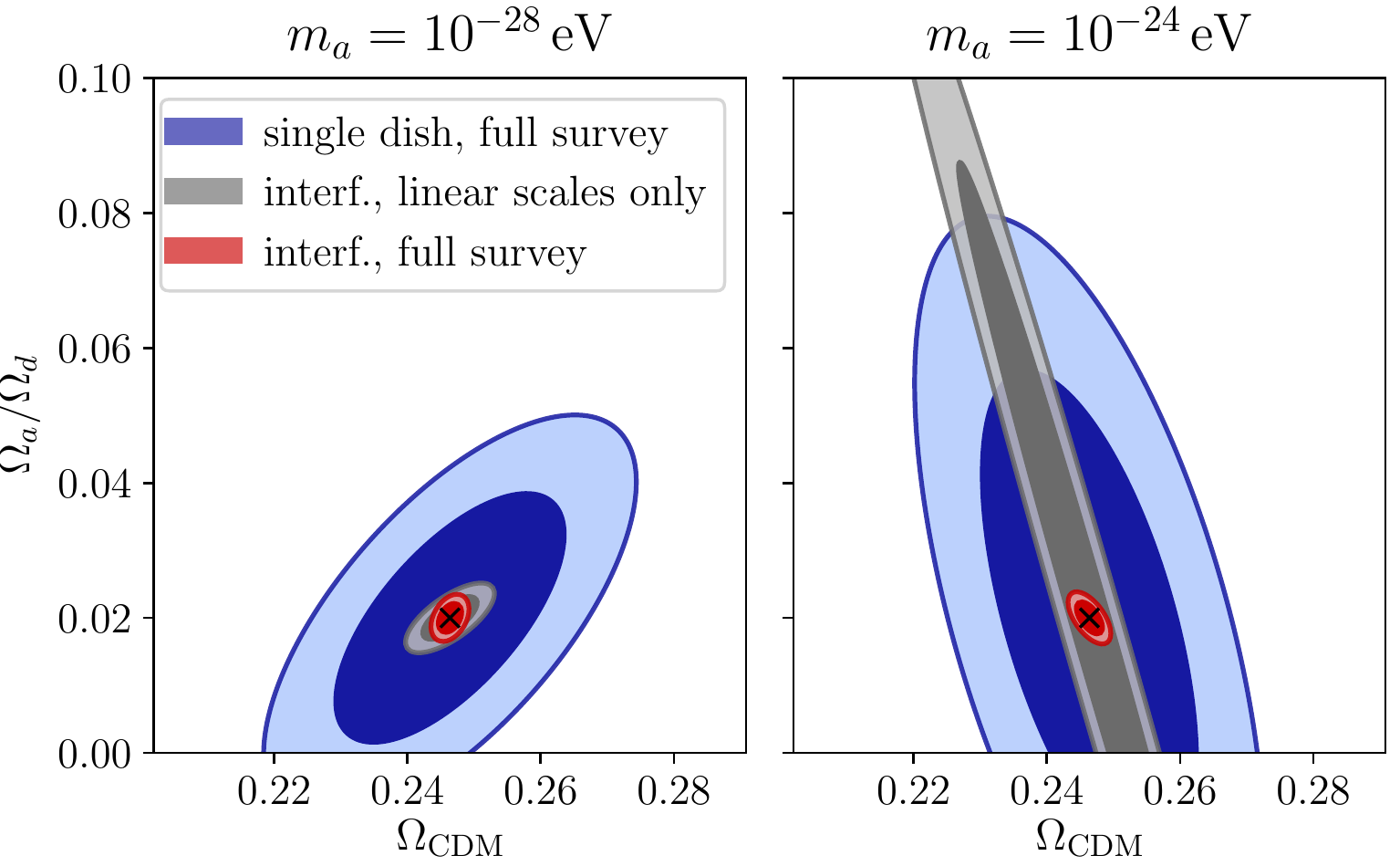}
    \caption{Error ellipses for $m_a = 10^{-28}$ (left-hand panel) and $m_a = \SI{e-24}{eV}$ (right-hand panel), obtained from the cumulative Fisher matrix for the interferometric SKA1MID configuration \citep[and the single-dish mode using the noise expression adapted from][]{Knox1995}.}
    \label{fig:CDM_error_ellipses}
\end{figure}
Fig. \ref{fig:CDM_error_ellipses} shows the error ellipses for two different masses and surveys: in the left-hand panel, for a mass of $m_a = \SI{e-28}{eV}$, and in the right-hand panel, a high mass of $m_a = \SI{e-24}{eV}$ for SKA1MID run in the interferometric and in the single-dish mode.
SKA1MID in the single-dish mode probes smaller $\ell$ and, thus, larger scales. 
SKA1MID in the interferometric mode, on the other hand, probes smaller, mostly non-linear scales.
The error ellipses for a mass $m_a = \SI{e-24}{eV}$ show that, indeed, a degeneracy if run in the single-dish mode and in the interferometric mode when only linear scales are included. 
For the full interferometric SKA1MID survey, the overall error on the axion fraction reduces significantly because degeneracies are broken with other parameters as mentioned above and  additional scales are included to the Fisher analysis.

For an axion of mass $m_a = \SI{e-28}{eV}$, a degeneracy with CDM is also observed, but in the opposite direction. 
This can be understood from the \HI power spectra in Fig. \ref{fig:P_HI}. For $m_a = \SI{e-28}{eV}$ on the range of scales where there is good signal-to-noise ratio, the axion just suppresses power, so it has the opposite degeneracy with CDM as for the heavier case.
This suppression is due to the suppression of matter fluctuations on already linear scales and relates to the lower Jeans wavenumber compared to higher axion masses. 
The degeneracy is less prominent for the full interferometric SKA1MID survey.

\subsection{Constraints on the Axion Fraction}
This section presents exclusion limits of the axion fraction by the IM surveys.
They are estimated by 
the minimum axion fraction that one is able to resolve with its $n \sigma$ marginal error. 
Explicitly, the $n \sigma$-exclusion limit on $\Omega_a/\Omega_d$ is obtained by the highest fiducial value of $\Omega_a/\Omega_d$ satisfying $n \sigma(\Omega_a/\Omega_d) = \Omega_a/\Omega_d$.

Furthermore, we combined the Fisher matrix from the IM surveys with those of future CMB measurements, namely from the Simons Observatory \citep{Ade2019_SO}, obtained from \citet{Hlozek_Marsh_forecast}, where ULAs were included in the analysis. 
This breaks the degeneracy between $A_s$ and the astrophysical parameters.
When combining CMB and IM surveys, it was necessary for most axion masses to extrapolate the marginal error to obtain the exclusion limits. 
However, the constraints are roughly constant, independent of the fiducial axion fraction, such that these exclusion limits should not be affected dominantly by the extrapolation. 

\begin{figure}
\centering
     \includegraphics[trim={1.5cm 0.6cm 1.5cm 0.75cm},clip,width=1.\linewidth]{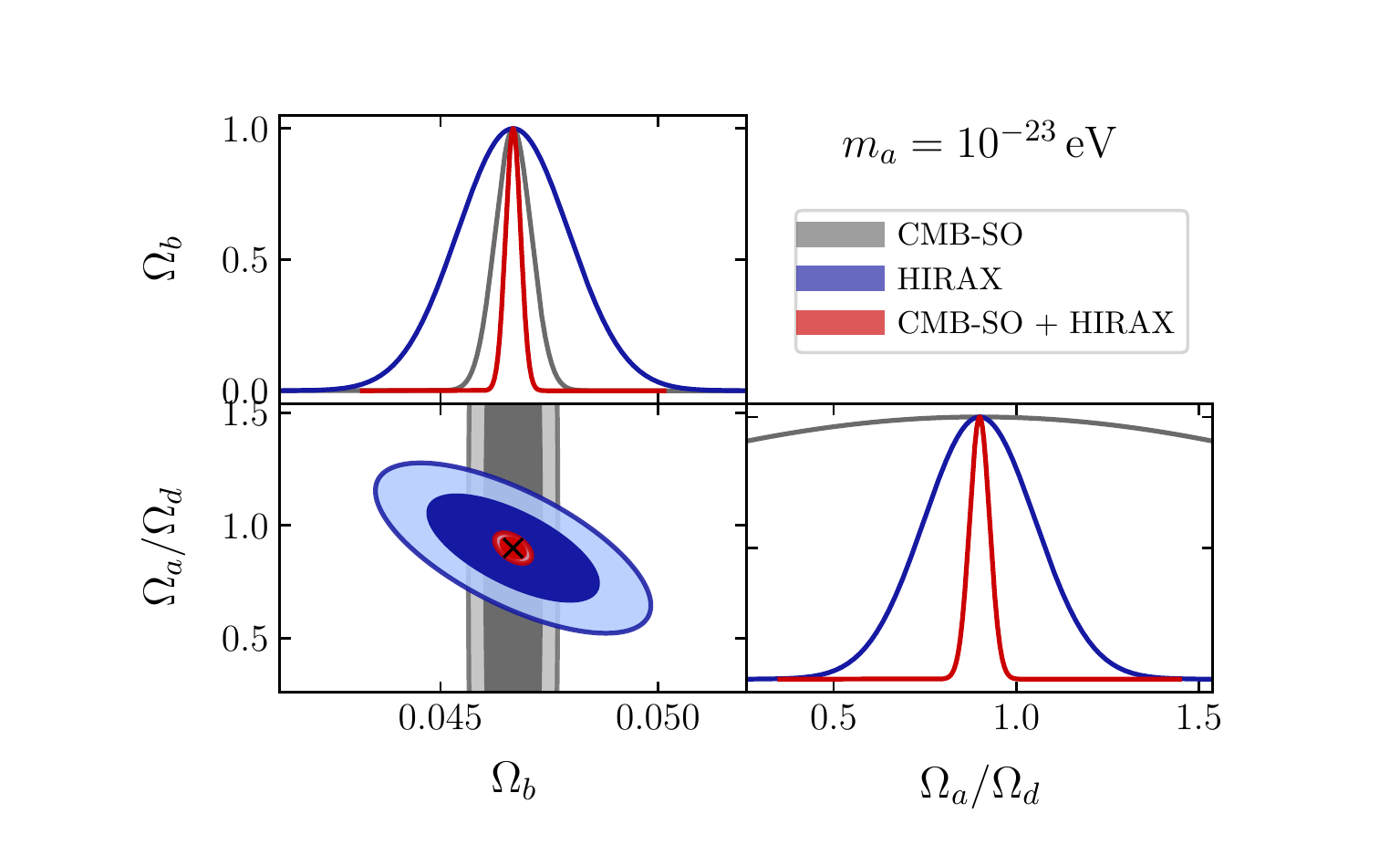}
    \caption{Error ellipses of the cosmological parameters from the HIRAX, CMB-SO surveys, and when combined. Fiducial parameters were used as in Table \ref{tab:parameters} with $m_a = \SI{e-23}{eV}$ and fiducial axion fraction $\Omega_a/\Omega_d = 0.9$.}
    \label{fig:error_ellipses_triangle_CMB+HIRAX}
\end{figure}

Fig. \ref{fig:error_ellipses_triangle_CMB+HIRAX} shows the effect of combining an IM survey with the CMB.
Upon combination of the surveys a marked decrease (by a factor of $\sim 10$) on the fractional axion density is observed for axions of mass below $\SI{e-24}{eV}$ for the HIRAX survey. 
CMB-SO and HIRAX surveys are highly complementary for this configuration: 
While the CMB Fisher matrix strongly constrains the cosmological parameters other than the axion fraction, the HIRAX survey is able to probe the axion fraction to greater accuracy.
Combining both Fisher matrices, therefore, results in a marked reduction of the marginal error on $\Omega_a/\Omega_d$.
This is exemplary shown for $\Omega_b$ and $\Omega_a/\Omega_d$ in Fig. \ref{fig:error_ellipses_triangle_CMB+HIRAX} for $m_a = \SI{e-23}{eV}$ and fiducial axion fraction $\Omega_a/\Omega_d = 0.9$. 
The CMB provides a precision measurement of $\Omega_b$, breaking the axion-baryon degeneracy in the HIRAX IM survey and significantly tightening the IM constraint on axion fraction, despite the CMB being insensitive to the axion fraction at this mass.

\begin{figure}
\centering
    \includegraphics[width=\linewidth]{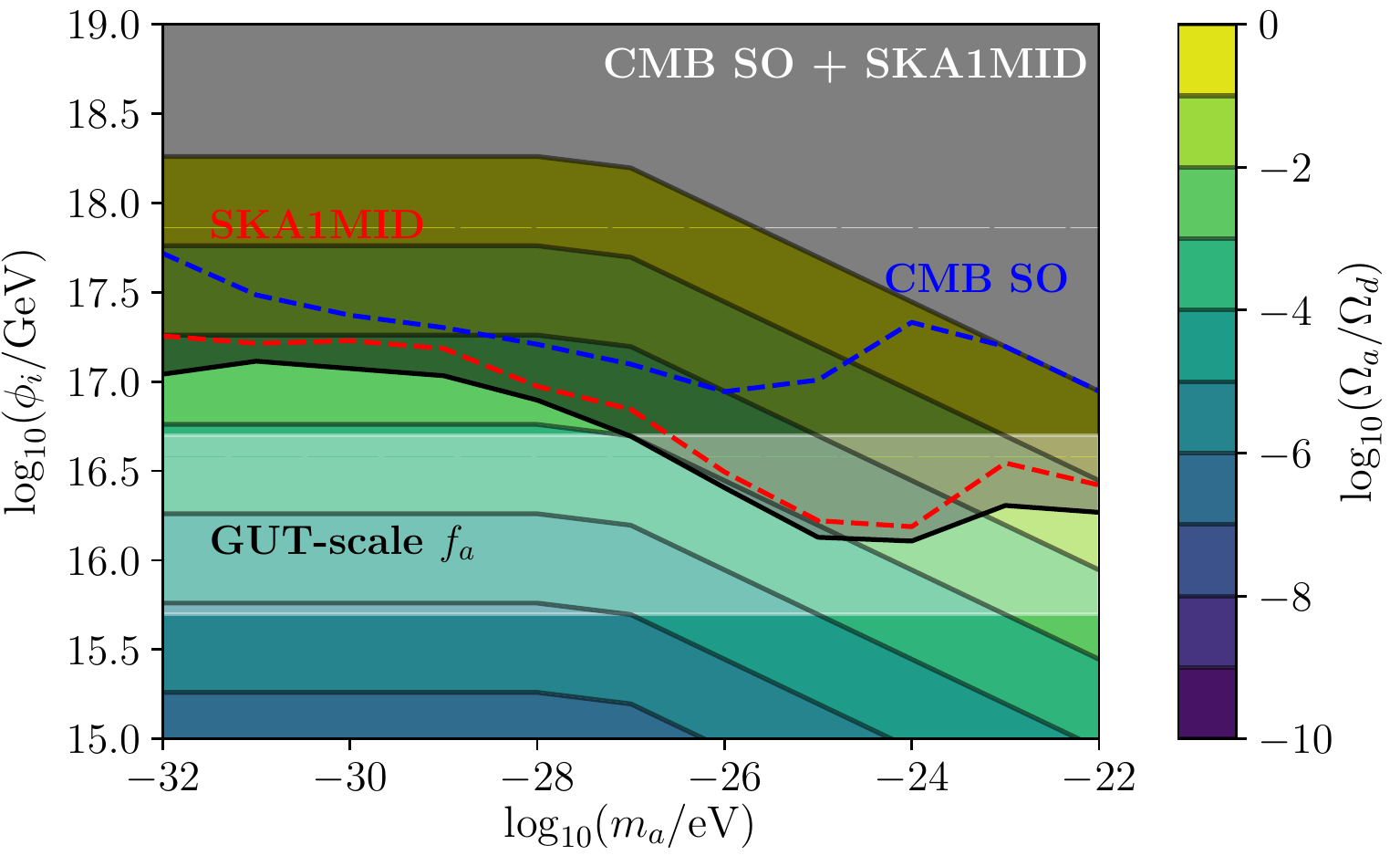}
    \caption{The expected axion fraction $\Omega_a/\Omega_d$ as a function of axion mass $m_a$ and initial field value $\phi_i$ according to equation \eqref{eq:axion_abundance}. 
The hatched, grey region shows the forecasted $2\sigma$ exclusion limits from SKA1MID (B1 + B2) run in the interferometric mode and the CMB-SO, while the blue and red line indicate those for CMB-SO and SKA1MID alone, respectively. The white shaded band shows the expectation for the GUT scale ($\sim \SI{e16}{GeV}$).}
    \label{fig:SKA_constraints}
\end{figure}
\begin{figure*}
\centering
    \includegraphics[width=\linewidth]{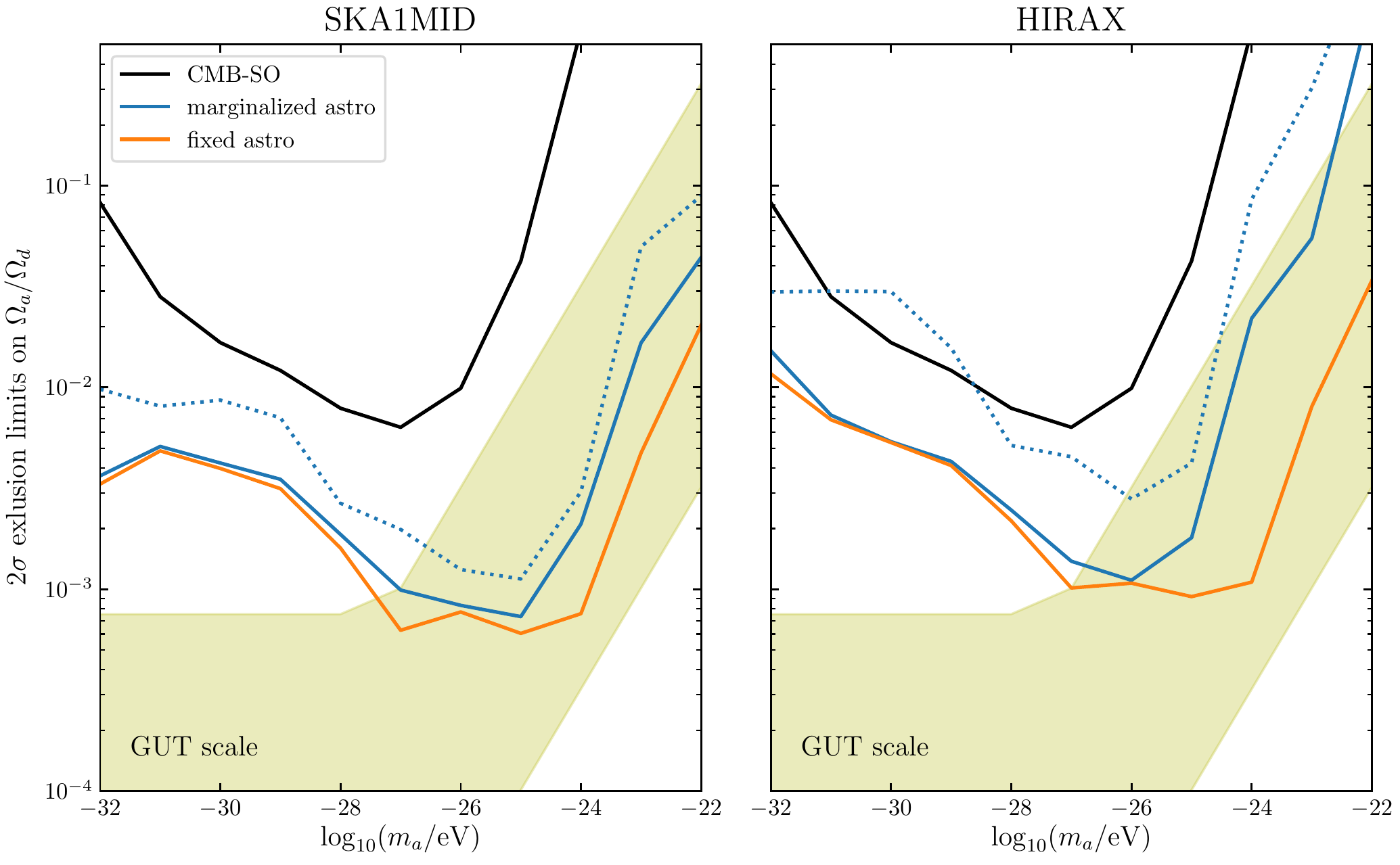}
    \caption{$2\sigma$ exclusion limits for the HIRAX and SKA1MID survey run in the interferometric mode. Dotted lines indicate the IM survey alone, while solid ones are for those combined with the forecasted CMB-SO constraints. The shaded regions indicate the expectation from the GUT scale. Depending on the precision of the astrophysical parameters, SKA1MID and HIRAX probe the interesting region near the GUT scale for ULAs with $m_a \gtrsim \SI{e-27}{eV}$. While axions with $m_a \lesssim \SI{e-25}{eV}$ leave a salient, scale-dependent imprint on linear scales due to their large de Broglie wavelength, heavier ones are degenerate with the astrophysical parameters on those scales. Breaking this degeneracy is possible by including smaller non-linear scales or by a more precise knowledge of the astrophysical parameters. With SKA1MID being more sensitive to smaller, non-linear scales than HIRAX, it is able to constrain the heavy fuzzy DM axions more strongly.}
    \label{fig:final_exclusion_limits}
\end{figure*}

Since degeneracies are broken between $\Omega_a/\Omega_d$ and the astrophysical parameters upon inclusion of non-linear scales, the SKA1MID survey is highly constraining. 
The obtained $2\,\sigma$-exclusion limits for this survey in the interferometric mode, the CMB-SO survey and when combined are shown in Fig. \ref{fig:SKA_constraints}, expressed in terms of the axion initial field value, which is related to the scale of spontaneous symmetry breaking, $f_a$.
\footnote{To relate the initial field value $\phi_i$ to $\Omega_a$, we neglected the error on $\Omega_d$ and correlations between these two, since the error on $\Omega_d$ is generally much smaller than for the axion density parameter (and we expect corrections by it to be subdominant given the overall uncertainty upon extrapolating the results).}
Excitingly, these forecasts within the present framework show that it is possible to probe the axion fraction to high precision for most masses. 
Even more, for intermediate and large axion masses, the exclusion limits are close to the GUT scale expectations. 
Fig. \ref{fig:final_exclusion_limits} summarizes the findings for the SKA1MID and the HIRAX surveys.

It should be possible to further reduce the exclusion limits by fixing the astrophysical parameters. 
This assumption is supported by the upcoming of current Ly\,$\alpha$ surveys (and future data releases), such as HETDEX \citep{HETDEX_2008}. These surveys should, in principle, be able to constrain the $M_\hi(M)$ relation and therefore astrophysical parameters further.
The results upon fixing the $M_{\hi}(M)$ relation for the HIRAX and SKA1MID survey are shown in Fig. \ref{fig:final_exclusion_limits}. 
As expected, the constraints from the HIRAX survey improve slightly for ULAs with the scale-dependent imprint and significantly for heavier ULAs where the influence in the bias are highly degenerate with the astrophysical parameters. 
Overall, probing the axion fraction on the sub-per-cent level is possible and the region near the GUT scale can be tested for axion masses $\gtrsim \SI{e-27}{eV}$.

\section{Discussion and Conclusions}
\label{sec:discussion}
In this paper, we have investigated the imprint of ULAs on late-time 21\,cm IM surveys in a mixed DM scenario.  
To do so we exploited the accurate, data-driven halo model introduced by \citet{Padmanabhan_etal2015}.
Axions were accommodated to this model by reference to massive neutrinos. 
Both were compared by looking at their variances. 
Thereby, a critical axion mass of  $\SI{e-27}{eV}$ was identified below which ULAs are treated similar to neutrinos (\HI as a tracer of the CDM and baryon field, but not the total matter field) and above which they are incorporated as a CDM-like component. 
The halo model was roughly checked against numerical results for heavier ULAs from \citet{Carucci_2017}. 
The present model adequately captures their main findings, providing further confidence for the proposed framework.
In contrast to the earlier work by \citet{Carucci_2017}, this study allows the axion fraction to be a free parameter (which is, to our knowledge, not studied in the literature at present).
This is important because it is predicted that these ULAs depending on their mass occur in subdominant abundances related to the GUT scale.

Heavier ULAs, $m_a \gtrsim \SI{e-24}{eV}$, suppress the formation of small-mass halos below the Jeans mass. Assuming a fixed amount of neutral hydrogen, more \HI needs to reside in heavier halos that are more strongly biased too than their low-mass counterparts. This leads to an increase in the \HI bias compared to $\Lambda$CDM. 
Note that Fig. \ref{fig:quant_comp_carucci} shows that at least some knowledge on the cut-off mass in the \HI--halo mass relation is necessary to constrain those ULAs: If the cut-off mass is larger than the Jeans mass of the ULA, the suppression of halos below the Jeans mass does not leave a ``fingerprint''.

For lighter ULAs, $m_a \sim \SI{e-28}{eV}$, an increase in the \HI bias is also observed, albeit out of different reasons: These ULAs suppress structure on almost all relevant scales (when compared to CDM), which makes their influence DE-like. The halo bias is increased for most halo masses and intermediate sized halos are suppressed. 

For ULAs with $m_a \lesssim \SI{e-25}{eV}$, the suppression in the power spectrum starting at scale $k_m$ and saturating at the Jeans scale $k_J$ becomes relevant even at large, linear scales in the \HI power spectrum (cf. equation \ref{eq:2-halo_term}), leading to a salient scale-dependent imprint of those ULAs.
For heavier ULAs the increase in the \HI bias is the main imprint on large, linear scales. The scale-dependence is partly ``hidden'': The suppression in the matter power spectrum happens at non-linear scales and for $m_a \gtrsim \SI{e-23}{eV}$ at scales where the one-halo term is dominant.

The scale-independent impact for ULAs of $m_a \gtrsim \SI{e-24}{eV}$ is degenerate with the astrophysical parameters, $v_{c,0}$ and $\beta$, which effectively control the \HI bias, and $A_s$. 
When only probing large, linear scales at a single redshift, this degeneracy cannot be broken.
Disentangling these parameters should be possible by (i) providing priors from other observations, (ii) including RSDs, (iii) probing non-linear scales, or (iv) the different redshift scalings of the impact of each parameter (cf. Section \ref{sec:degeneracy_structure}).

Forecasts for different surveys were run with help of a Fisher matrix analysis. Degeneracies were partly broken by priors (CMB or fixing astrophysics), the different redshift scaling of the impact of different parameters and for HIRAX only mildly probing non-linear scales (cf. Fig. \ref{fig:survey_scan_linear}).
The results give an exciting and promising impression: 
within this framework combined future CMB and IM observations should be able to test axion fractions to the per cent level or even below.
It is possible to probe the interesting region near the GUT scale in the mass region where the ULA imprint the scale-dependent feature onto the 21\,cm signal.
Including non-linear scales and combining a SKA1MID-like IM survey with the Simons Observatory CMB, the benchmark fuzzy DM model with $m_a = \SI{e-22}{eV}$ can be constrained at few per cent. This is almost an order of magnitude improvement over current limits from the Ly\,$\alpha$ forest \citep{Kobayashi_etal2017}.
For the HIRAX survey, which probes only mildly the non-linear regime, astrophysics need to be known precisely such that current and future constraints could be greatly improved for fuzzy DM. 

These exciting results call for other studies to check the robustness of the present results, which includes calibrating the halo model in the mixed ULA-CDM scenario. 
The following list includes the most important assumptions with respect to \HI IM, which could be checked in more detail and give guidance to future studies:
\begin{enumerate}
\item fixed and redshift-independent $\Omega_\hi$;
\item no foreground contamination;
\item cosmological information from $b_\hi$;
\item no inclusion of RSDs;
\item no specific impact of axions onto the neutral hydrogen profile;
\item use of Limber approximation.
\end{enumerate}
First, given the current poor constraints on $\Omega_\hi$, this is an optimistic assumption at the present day.
Also, \HI IM surveys alone measuring the large, linear scales can only estimate the quantity $\Omega_\hi b_\hi$. 
Because the main effect of ULAs with $m_a \gtrsim \SI{e-24}{eV}$ is the increase in \HI bias, it is necessary to know the precise value of $\Omega_\hi$.
A possible resolution is to determine $\Omega_\hi$ with other observations or
to break this degeneracy by the inclusion of RSDs into our analysis.
The latter requires a model of RSDs which includes axions and also might provide by itself additional information on the axion fraction.
Hence, the inclusion of RSDs to the analysis is an important extension.

Secondly, foregrounds have been neglected and a perfect foreground removal has been assumed.
Given that foregrounds are expected to be the leading systematic factor for \HI IM surveys, we discussed the influence of foreground residuals on the angular power spectrum briefly in Section \ref{sec:foregrounds}. We concluded that efficient foreground removal is necessary and a more rigorous treatment (with the inclusion of the foreground wedge) within the forecasts is a relevant next step.

Thirdly, the information from modelling the bias parameter have been included in the analysis.
This touches a central point for any large-scale structure survey: it is necessary to have a sufficient knowledge on the exact behavior and modelling of the bias parameter to rely on the results.
If not, deviations from an expected ($\Lambda$CDM) signal of LSS survey could be alleviated by an inaccurate modelling of the bias. 
This works also the other way round: A deviation from $\Lambda$CDM could be hidden by an inaccurate model of the bias, giving rise to a potential psychological \textit{confirmation bias}. 
Two of such potential pitfalls when modelling the bias shall be stated. First, it could be expected that the neutrino-like ULAs introduce a slight scale dependence to the bias as simulations show for massive neutrinos \citep{cosmo_neutrinosI}, which the present model does not capture.
Note, however, that this effect is reduced by our choice to only consider the CDM and baryon field.
Secondly, to what extent it can be expected that the bias parameter is scale-independent on large scales even for the $\Lambda$CDM case, is an ongoing question and large simulations
 are employed to investigate the \HI bias relation further \citep{ingredients_for_21cm_IM}. 
To give an idea on the needed accuracy in the modelling of the \HI bias in the present case for the heavier ULAs ($m_a \gtrsim \SI{e-24}{eV}$), 
Fig. \ref{fig:b_HI} gives a rough estimate: The \HI bias changes at $\sim 10\%$ level for axion fractions at $m_a = \SI{e-24}{eV}$ on the $\sim 1\%$. 
Thus, precision in \HI bias modelling needs to be below the $10\%$ ballpark for constraints on the per cent level (and in that mass region) to be robust.
A more conservative approach could relax the assumption of precise modelling knowledge of the \HI bias, e.g. by introducing nuisance parameters in the redshift dependence.

Also, this model is partly ignorant of the influence of axions to the exact \HI radial profile. 
Some clear deviations from pure CDM could be expected \citep{Veltmaat_etal2019} and modeled more accurately. This would also provide additional information on the axion fraction, although the inclusion of highly non-linear scales is necessary.
Apart from that, the effect of the Limber approximation could be checked against more accurate calculations of the angular power spectrum. This study could also include a thorough investigation of the effect of (un)equal-time correlators (which might become important). 

In short, all of the above points are of importance to adequatly capture future 21\,cm IM data and while relaxing the assumptions (i) to (iii) will likely weaken the present constraints, the opposite will be true for a more accurate modelling of (iv) and (v). 
An important future step is to investigate the HMF and halo bias in the mixed CDM-axion DM scenario with simulations and calibrating the model.
We leave these extensions and simulations to future work.

%%% ------------------------ ACKNOWLEDGEMENTS ----------------------------------
\section*{Acknowledgments}
We thank Christoph Behrens for fruitful discussions.
This work made use of the open source packages \textsc{numpy} and \textsc{scipy}~\citep{numpy, scipy}. JBB and DJEM are supported by the Alexander von Humboldt Foundation and the German Federal Ministry of Education and Research. 
RH is a CIFAR Azrieli Global Scholar, Gravity and the Extreme Universe Program, 2019, and a 2020 Alfred. P. Sloan Research Fellow. RH is supported by Natural Sciences and Engineering Research Council of Canada. The Dunlap Institute is funded through an endowment established by the David Dunlap family and the University of Toronto. RH acknowledges that the land on which the University of Toronto is built is the traditional territory of the Haudenosaunee, and most recently, the territory of the Mississaugas of the New Credit First Nation. She is grateful to have the opportunity to work in the community, on this territory.
HP acknowledges support from the Swiss National Science Foundation under the Ambizione grant PZ00P2\_179934.

\section*{Data Availability}
The data underlying this study are available in this paper. The derived data in this research can be generated with the code published along this paper (\url{https://github.com/JurekBauer/axion21cmIM.git}) and will also be shared on reasonable request to the corresponding author.
%%%%%%%%%%%%%%%%%%%%%%%%%%%%%%%%%%%%%%%%%%%%
%% -------------------------- APPENDIX ------------------------------------------%%%%%%
%%%%%%%%%%%%%%%%%%%%%%%%%%%%%%%%%%%%%%%%%%%

\newpage

\appendix

\section{Comparison to single-dish noise given in Bull et al. (2015)}
\label{sec:appendix}
In appendix D of \citet{Bull_Ferreira2015}, a formula for the temperature noise is given by the Gaussian root mean square width
\begin{align*}
\sigma_T \approx \frac{T_{\text{sys}}}{\sqrt{n_{\text{pol}} \Delta \nu t_{\text{obs}}}} \frac{\lambda^2}{\theta_B^2 A_e} \sqrt{\Omega_{\text{surv}}/\theta_B^2} \sqrt{\frac{1}{N_d N_b}}.
\end{align*}

If one ignores the beam responses for the moment, the 3D power spectrum is given by $P_N = \sigma_T^2 V_{\text{pix}}$, where $V_{\text{pix}} = (r\theta_B)^2 \times (r_\nu \delta \nu / \nu_{21})$ is the 3D volume of each volume element with $r_\nu = c (1+z)^2/H(z)$.

To convert $P_N$ into $N_{\ell}$, we take 
\begin{align}
\nonumber
&N_{\ell}(z_i,z_j) = \frac{2}{\pi} \int \diff z\,W_i(z) \int \diff z'\,W_{j}(z') \times \\
&\qquad \qquad \int \diff k\,k^2 P_N(k, z, z') j_\ell(kr(z)) j_\ell(kr(z')),
\label{eq:Nell_from_PN}
\end{align}
where $W_i$ and $W_{j}$ are the window function for the redshift bins $z_i$ and $z_j$, $j_{\ell}$ denotes the spherical Bessel function of rank $\ell$, and $r(z)$ is the comoving distance. 

The expression above in equation \eqref{eq:Nell_from_PN} can be simplified upon assuming the instrumental noise power spectrum is $k$-independent:
\begin{align*}
\nonumber
N_\ell &= \int {\text{d}}z \frac{H(z)}{c} \frac{W_i^2(z)}{r^2(z)} P_N(z_i) \\ 
\nonumber &\approx \Delta z W^2(z) \frac{H(z)}{c r^2(z)} V_{\text{pix}} \sigma_T^2 \\ 
\nonumber &\approx \theta_B^2 \sigma_T^2 
\end{align*}

This yields
\begin{align*}
N_\ell  = \frac{T_{\text{sys}}^2 \Omega_{\text{surv}}}{n_{\text{pol}} \Delta \nu t_{\text{obs}}} \frac{\lambda^4}{\theta_B^4 A_e^2} \frac{1}{N_d N_b}
\end{align*}

The beam frequency and angular responses is given by
\begin{align}
B^{-1} = B_\perp^{-2} B_\parallel^{-1},
\label{eq:Bull_beam-response}
\end{align}
with 
\begin{align*}
B_\parallel &= \exp\left[- \frac{(k_\parallel r_\nu \delta \nu/\nu_{21})^2}{16 \ln{2}}\right] \quad {\text{and}} \\
B_\perp &= \exp\left[- \frac{(k_\perp r \theta_B)^2}{16 \ln{2}}\right].
\end{align*}

Assuming $k_\parallel$ to be small, one can take $\ell + (1/2) = rk_\perp$ such that for large $\ell$,
\begin{align*}
B^{-1} \approx \exp\left[\frac{\ell^2 \theta_B^2}{8 \ln{2}}\right].
\end{align*}
We conclude for the dimensionless noise expression:
\begin{align}
N_\ell = \frac{T_{\text{sys}}^2 \Omega_{\text{surv}}}{\bar{T}_b^2 n_{\text{pol}} \Delta \nu t_{\text{tot}}} \frac{\lambda^4}{\theta_B^4 A_e^2} \frac{1}{N_d N_b} \exp\left[\frac{\ell^2 \theta_B^2}{8 \ln{2}}\right].
\label{eq:Bull_full-noise-expression}
\end{align}

Upon redefining the number of beams in the upper expression to include the number of polarization modes, the difference between this expression and the expression adapted from \citet{Knox1995} in equation \eqref{eq:Knox_noise} resides in the inclusion of the effective area of the dish. 
This results in a different prefactor of $\sim 3$.
\newpage
  
\bibliographystyle{mnras}
\bibliography{refs}
\end{document}